\documentclass[twocolumn,showpacs,amsmath,amssymb,prd,floatfix]{revtex4}
\usepackage{graphicx}% Include figure files
\usepackage{dcolumn}% Align table columns on decimal point
\usepackage[varg]{txfonts}
\usepackage{bm}% bold math

\begin{document}
%\begin{CJK*}{UTF8}{min} % Use default fonts from CJK (see below)

\title{
  Nonlinear Perturbation Theory Integrated with Nonlocal Bias,
  Redshift-space Distortions, and Primordial Non-Gaussianity
}

%\author{Takahiko Matsubara (松原隆彦)}
\author{Takahiko Matsubara}
\email{taka@a.phys.nagoya-u.ac.jp}
\affiliation{%
  Kobayashi-Maskawa Institute for the Origin of Particles and the
  Universe, Nagoya University, Chikusa, Nagoya, 464-8602, Japan; }%
\affiliation{%
  Department of Physics, Nagoya University, Chikusa, Nagoya, 464-8602,
  Japan}%

\date{\today}% It is always \today, today,

\begin{abstract}
    The standard nonlinear perturbation theory of the gravitational
    instability is extended to incorporate the nonlocal bias,
    redshift-space distortions, and primordial non-Gaussianity. We
    show that local Eulerian bias is not generally compatible to local
    Lagrangian bias in nonlinear regime. The Eulerian and Lagrangian
    biases are nonlocally related order by order in the general
    perturbation theory. The relation between Eulerian and Lagrangian
    kernels of density perturbations with biasing are derived. The
    effects of primordial non-Gaussianity and redshift-space
    distortions are also incorporated in our general formalism, and
    diagrammatic methods are introduced. Vertex resummations of
    higher-order perturbations in the presence of bias are considered.
    Resummations of Lagrangian bias are shown to be essential to
    handle biasing schemes in a general framework.
\end{abstract}

\maketitle
%\end{CJK*}

\section{\label{sec:intro}
Introduction
}

The large-scale structure of the universe is one of the most powerful
probes in cosmology. The statistical nature of primordial density
fluctuations can be investigated by large-scale distributions of
galaxies. Geometrical effects on the observed clustering pattern of
galaxies can constrain the nature of dark energy, through the
Alcock-Paczinski effect \cite{AP79,BPH96,MS96} or baryon acoustic
oscillations (BAO) \cite{EHT98,mat04,eis05}. Recently, it is found
that the presence of primordial non-Gaussianity introduces
scale-dependent bias in the halo clustering
\cite{dal08,MV08,slo08,TKM08,DS09}. Therefore, many possible models in
cosmology, such as inflationary scenarios, dark energy models,
modified gravity, and so on, should be constrained by precision
measurements of the large-scale structure in near future.

To compare the observations with theory, it is crucial to make precise
predictions of observable quantities from a given cosmological model.
The linear theory applies on very large scales \cite{pee80}. However,
the linear theory is not sufficiently accurate for purposes in the
precision cosmology.

Accurate predictions of statistical measures of galaxy clustering
beyond the linear theory are provided by nonlinear theories. The
method of numerical simulation is one of the most straightforward ways
of investigating nonlinear dynamics. However, they are not free from
numerical artefacts and systematics, such as finite-volume effects,
finite-resolution effects, and so forth. Fortunately, the analytical
perturbation theory is applicable on large scales where density
fluctuations are small. Thus the nonlinear perturbation theory of
gravitational instability attracts renewed interests in recent years.

The nonlinear perturbation theory have been developed since decades
ago \cite{jus81,vis83,fry84,GGRW86,MSS91,JB94,SF96,BCGS02}. The
traditional perturbation theory is formulated in Eulerian space, and
such theory is referred to as the standard perturbation theory (SPT).
The perturbation theory in Lagrangian space is also formulated
\cite{buc89,mou92,buc92,BE93,buc94,HBCJ95,cat95,CT96,EB97}, which is
called the Lagrangian perturbation theory (LPT). The first-order LPT
corresponds to the classic Zel'dovich approximation \cite{zel70}.

In these years, the renormalized perturbation theory (RPT)
\cite{CS06a,CS06b} and other approaches
\cite{mcd07,val07,MP07,MP08,TH08,pie08,mat08a} have been developed to
improve the accuracy of perturbation theory in nonlinear regime,
partially taking into account higher-order effects of the SPT. Some of
those approaches are based on the reformulation of fluid equations
using the propagator, the vertex, and a source \cite{sco01}, which
provides a way to use standard tools of field theory. Nevertheless,
various levels of approximations and ansatz should be employed in
those approaches.

The RPT and its variants mentioned above significantly improve the
perturbation theory of dark matter in real space. However, one of the
most important applications of the perturbation theory is to interpret
the large-scale clustering of galaxies or other astronomical objects,
observed by redshift surveys. The observable quantity in redshift
surveys is the distribution of objects in redshift space. Even though
the RPT and its variants could be powerful in predicting the nonlinear
power spectrum of dark matter in real space, one could not directly
compare the theoretical prediction with observations.

There are two obstacles to the comparison between the improved
perturbation theories and observations. The first one is the
redshift-space distortions: the redshift as a measure of the radial
distance is contaminated by peculiar velocities. It is straightforward
to take them into account in the SPT framework
\cite{kai87,ham92,HMV98,SCF99}. However, the SPT in redshift space
breaks down at larger scales than in real space and the applicability
range of scales is fairly narrow \cite{SCF99}, since the SPT does not
sufficiently reproduce the nonlinear smearing effects, known as the
Fingers-of-God (FoG) effect \cite{jac72,ST77}.

Nonlinear modelings of the redshift-space distortions beyond the SPT
are proposed \cite{PD94,sco04,tar10}, in which the FoG effects are
phenomenologically put by hand. It is found in those studies that the
FoG effects can be represented by putting a Gaussian damping factor in
front of the power spectrum. It is shown that the Gaussian damping
factor in redshift space is naturally derived from the LPT
\cite{mat08a}, where the phenomenological Gaussian factor should be
modified and additional mode-coupling terms should be taken into
account in nonlinear redshift space.

The second obstacle to the comparison between perturbation theories
and observations is the biasing. Any astronomical objects which can be
observed are biased tracers of underlying mass distributions. In the
galaxy redshift surveys, the tracers are galaxies. The exact
relationship between the distribution of mass and that of galaxies
depends on the complex, nonlinear process of galaxy formation which is
not clearly understood.

Analytic models of biasing have been proposed, including the model of
local Eulerian bias \cite{col93,FG93,SW98}, the halo model
\cite{MW96,MJW97,ST99,SSHJ01,CS02}, peaks model
\cite{kai84,DEFW85,BBKS,RS95,des08,DS10,DCSS10}, etc. The first model
is a purely phenomenological parameterization of the bias, assuming a
local relation between the mass and galaxy distributions. The last two
models are relatively more physical than the first one, and are
categorized as the Lagrangian bias, i.e., the locations of the galaxy
formation are specified in initial density fields. The location of a
peak, for example, is displaced by the dynamical evolution of density
fluctuations. In the framework of SPT, the local Eulerian bias is
usually adopted \cite{HMV98,SCF99,tar00,mcd06,JK08}. However, as shown
below in this paper, the halo model and the peaks model are not
compatible with the local Eulerian bias in nonlinear regime, because
the gravitational evolutions are nonlocal process in general.

\begin{figure}
\begin{center}
\includegraphics[width=18pc]{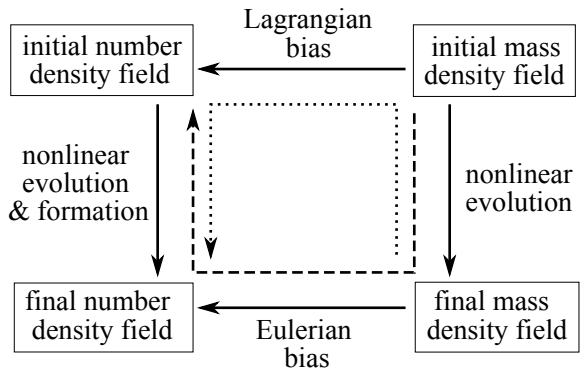}%{biases.eps}
\caption{\label{fig:Biases} The relation between Eulerian and
  Lagrangian biases. The Eulerian bias is expressible by the
  Lagrangian bias (dotted arrow) and vice versa (dashed arrow), only
  when the biases are allowed to be nonlocal. Note that nonlinear
  evolutions and formation of objects are nonlocal processes in
  nonlinear regime. }
\end{center}
\end{figure}

In Fig.~\ref{fig:Biases}, the relation between the Eulerian and
Lagrangian biases is shown.
The final mass density field is the result of nonlinear evolutions
from the initial mass density field. The final number density field of
objects, which are observable such as galaxies, is the result of
nonlinear evolutions and formations of those objects. The sites of the
formation in Lagrangian space define the initial number density field.
Thus the initial number density field also depends on the redshift of
observation. The Lagrangian bias corresponds to the relation between
the initial mass density field and the initial number density field,
while the Eulerian bias corresponds to the relation between the final
mass density field and the final number density field.

The initial number density field is constructed only when the
formation process is known. For example, the halo model and the peaks
model give prescriptions of constructing the initial number density
field from the initial mass density field; i.e., these models specify
the Lagrangian bias. In the halo model with the peak-background split
\cite{MW96}, the initial number density is locally determined by
initial mass density field. Thus the halo model corresponds to a local
Lagrangian bias. In the peaks model, the sites of the formation is not
locally determined, since the peaks are defined not only by local
values, but also by spatial derivatives of the field \cite{BBKS}. Thus
the peaks model corresponds to a (semi-)nonlocal Lagrangian bias. Such
models of Lagrangian bias are more physically motivated than the
Eulerian bias. The bias depends on the properties of observed objects,
such as the mass and luminosity. The Lagrangian bias schemes described
above actually depend on the mass (or the peak hight) of collapsed
objects.

The local Lagrangian bias and the local Eulerian bias are not
compatible to each other in nonlinear regime. This fact is well
illustrated in Fig.~\ref{fig:Biases}. The Eulerian bias is expressible
by the Lagrangian bias (dashed arrow) and vice versa (dotted arrow).
The relations involves nonlinear evolutions of mass, and nonlinear
evolutions and formation of observed objects. Nonlinear evolutions and
formations are nonlocal processes. Therefore, the Eulerian bias should
be nonlocal even when the Lagrangian bias is local. The local biasing
schemes are compatible only in the case that the linear theory and a
local approximation of formation process are valid. Such conditions do
not apply in generally nonlinear regime.

In reality, the bias is definitely nonlocal. For example, the galaxies
are largely affected by their environment \cite{ben10}. The
nonlocality should be taken into account in a precise modeling of the
bias. The exact forms of the nonlocality in formation processes of
observed objects have not been fully understood. It requires a lot of
future work with analytic, numerical, and observational studies of
complicated astrophysical processes to understand the exact
nonlocality of bias.

One of the main purposes of this paper is to formulate a nonlinear
perturbation theory which can handle both the Eulerian and Lagrangian
nonlocal biases in general. There mainly two merits in this
formulation. First, one can distinguish general properties of
nonlinear clustering which do not depend on details of formation
processes from those which largely depends on models of bias. Second,
it is straightforward to predict observable quantities in any models
of bias. Effects of redshift-space distortions and primordial
non-Gaussianity are also included in the formalism. Resummation
techniques in the presence of nonlocal bias are introduced as well.
The resummations of bias are shown to be essential to handle the
nonlocal bias in a general way.

This paper is organized as follows. In Sec.~II, both Eulerian and
Lagrangian perturbation theories in real space are extended to include
both the nonlocal bias and the primordial non-Gaussianity.
Diagrammatic methods with graphical representations are introduced.
The relation between the perturbative kernels of EPT and LPT with
nonlocal bias is derived as well. In Sec.~III, the formalism of the
previous section is extended to include the effect of redshift-space
distortions. Techniques of the vertex resummations in our formalism
are introduced in Sec.~IV, and finally, some models of the Lagrangian
bias are considered, illustrating how higher-order bias factors are
evaluated in our formalism.

\section{\label{sec:PTreal}
Nonlinear perturbation theory with Nonlocal bias
}

\subsection{\label{subsec:EPT}
Nonlocal bias in Eulerian space
}

In the following, the density contrast of observed objects X at a
comoving, Eulerian position $\bm{x}$ is denoted by $\delta_{\rm
  X}(\bm{x})$. The observed objects X can be any astronomical objects
such as galaxies, quasars, absorption lines, 21cm emissions, and so
forth, which are selected and catalogued in a given redshift survey.
The density field of the objects is not generally a local nor linear
function of the underlying mass density contrast $\delta_{\rm
  m}(\bm{x})$. Instead, they are nonlocally and nonlinearly related to
each other in general. In other words, the density contrast
$\delta_{\rm X}$ of objects is a functional of the mass density
contrast $\delta_{\rm m}$.

The functional relation between density fields of mass and objects is
deterministic on scales we are interested in. One might think that the
formation process of objects is determined not only by the density
field of mass, but also by other physical factors such as the local
radiation density and its spectrum, merger histories of galaxies, etc.
However, the dynamical evolutions in the structure formation are
deterministic, and the initial density field uniquely determines all
the subsequent states of the universe, including the above complex
factors. Consequently, when only the growing mode solutions are kept,
density contrasts of mass and objects are nonlocal functionals of
initial density contrast $\delta_{\rm L}$: we have $\delta_{\rm m} =
{\cal F}_{\rm m}[\delta_{\rm L}]$ and $\delta_{\rm X} = {\cal F}_{\rm
  X}[\delta_{\rm L}]$, where ${\cal F}_{\rm m}$ and ${\cal F}_{\rm X}$
represent nonlocal functionals.

On scales where the perturbation theory is applicable, the motion of
dark matter is single streaming. In which case, the spatial
distribution of dark matter uniquely inverted to give the initial
density field $\delta_{\rm L} = {\cal F}_{\rm m}^{-1}[\delta_{\rm
  m}]$. Thus we have a deterministic functional relation $\delta_{\rm
  X}[\delta_{\rm m}] = {\cal F}_{\rm X}[{\cal F}_{\rm
  m}^{-1}[\delta_{\rm m}]]$. Taking into account the translational
invariance, the Taylor expansion of the functional is generally given
by
\begin{multline}
  \delta_{\rm X}(\bm{x}) =
  \sum_{n=0}^\infty \frac{1}{n!} \int d^3x_1\cdots d^3x_n\,
  b_n(\bm{x}-\bm{x}_1,\ldots,\bm{x}-\bm{x}_n) \\
  \times \delta_{\rm m}(\bm{x}_1) \cdots \delta_{\rm m}(\bm{x}_n).
\label{eq:1-1}
\end{multline}
The nonlocal bias functions, \{$b_n$\}, specify the relation between
the number density field of objects and mass density field. The first
term with $n=0$ in Eq.~(\ref{eq:1-1}) gives just a constant $b_0$,
which is determined by other functions $b_n$ to ensure a condition
$\langle\delta_{\rm X}\rangle=0$. The constant term $b_0$ is
irrelevant as we consider connected moments of the density contrast.
In Fourier space, moreover, the constant term just disappears when
nonzero modes $\bm{k}\ne \bm{0}$ are considered. Therefore, we do not
retain the constant term $b_0$ in the below.

The Taylor expansion of Eq.~(\ref{eq:1-1}) is applicable only when the
functional dependence is smooth. However, as shown in
Sec.~\ref{sec:VResum} below, the technique of vertex resummation
relaxes this constraint, and even singular dependences of the number
density field on the mass density field can be handled with this
technique.

The local biasing ansatz is recovered when
$b_n(\bm{x}_1,\ldots,\bm{x}_n)$ is replaced by $W_R(\bm{x}_1) \cdots
W_R(\bm{x}_n)\,b_n$, where $b_n$ is now a constant for each $n$, and
$W_R(\bm{x})$ is a smoothing kernel with a smoothing radius of $R$.
However, the Eulerian local biasing scheme is not a natural model when
we consider nonlinear dynamics, as we explicitly show below in this
paper.

In Fourier space, Eq.~(\ref{eq:1-1}) reduces to
\begin{multline}
  {\delta}_{\rm X}(\bm{k})
  = \sum_{n=1}^\infty \frac{1}{n!}
  \int \frac{d^3k_1}{(2\pi)^3} \cdots \frac{d^3k_n}{(2\pi)^3}
  (2\pi)^3 \delta_{\rm D}^3(\bm{k}_{1\cdots n} - \bm{k}) \\
%  (2\pi)^3 \delta_{\rm D}^3(\bm{k}_1+\cdots +\bm{k}_n - \bm{k}) \\
  \times {b}_n(\bm{k}_1,\ldots,\bm{k}_n)\, 
  {\delta}_{\rm m}(\bm{k}_1) \cdots {\delta}_{\rm m}(\bm{k}_n),
\label{eq:1-2}
\end{multline}
where we use a notation,
\begin{equation}
  \bm{k}_{1\cdots n} \equiv \bm{k}_1+\cdots +\bm{k}_n,
\label{eq:1-2-1}
\end{equation}
throughout this paper. Some variables like $\delta_{\rm m}$,
$\delta_{\rm X}$ and $b_n$ in Fourier space are denoted by the same
symbols as those in real space, instead of properly using symbols like
$\tilde{\delta}_{\rm m}$, $\tilde{\delta}_{\rm X}$, $\tilde{b}_n$,
etc. We will work in Fourier space in most of this paper. The
convention of Fourier transform and its inverse in this paper is given
by
\begin{equation}
  \tilde{F}(\bm{k}) = \int d^3x\, e^{-i\bm{k}\cdot\bm{x}} F(\bm{x}),\quad
  \tilde{F}(\bm{x}) = \int \frac{d^3k}{(2\pi)^3}
  e^{i\bm{k}\cdot\bm{x}} \tilde{F}(\bm{k}).
\label{eq:1-2-2}
\end{equation}

Since the bias relations should not depend on the coordinates system,
the bias functions $b_n$ should be rotationally invariant. For
example, the first-order bias function $b_1(\bm{k})$ is actually a
function of the magnitude $k=|\bm{k}|$ and can be denoted as $b_1(k)$.
Similarly, the second-order bias function $b_2(\bm{k}_1,\bm{k}_2)$ is
actually a function of $k_1$, $k_2$, and $k_{12} = |\bm{k}_1 +
\bm{k}_2|$ which characterize relative configuration of $\bm{k}_1$ and
$\bm{k}_2$. Thus the function can be denoted as $b_2(k_1,k_2;k_{12})$,
where the first two arguments are symmetric under a permutation, but
the last one. Similarly, the third-order bias function can be denoted
as $b_3(k_1,k_2,k_3;k_{12},k_{23},k_{31})$, and higher-order bias
functions depend only on rotationally invariant set of variables.

In the case of local biasing ansatz, $b_n(\bm{k}_1,\ldots,\bm{k}_n)$
is replaced by $W(k_1R)\cdots W(k_nR)\, b_n$, where $b_n$ is now a
constant for each $n$, and $W(kR)$ is a smoothing window function.
When the smoothing radius $R$ is much smaller than the clustering
scales we are interested in, the smoothing window function can be
dropped and each bias function is simply considered as a constant
$b_n$ in Fourier space.

The mass density contrast $\delta_{\rm m}(\bm{x})$ is also a nonlocal
and nonlinear functional of a linear density field $\delta_{\rm
  L}(\bm{x})$. We have a Taylor expansion which has a similar form
with Eq.~(\ref{eq:1-2}) in Fourier space:
\begin{multline}
  {\delta}_{\rm m}(\bm{k}) =
  \sum_{n=1}^\infty \frac{1}{n!}
  \int \frac{d^3k_1}{(2\pi)^3} \cdots \frac{d^3k_n}{(2\pi)^3}
  (2\pi)^3 \delta_{\rm D}^3(\bm{k}_{1\cdots n} - \bm{k}) \\
  \times F_n(\bm{k}_1,\ldots,\bm{k}_n) 
  {\delta}_{\rm L}(\bm{k}_1) \cdots {\delta}_{\rm L}(\bm{k}_n),
\label{eq:1-3}
\end{multline}
where $F_n$'s are perturbative kernels. We consider the density fields
at any given redshift, and time-dependences are suppressed in the
above notations. For the linear density field, $\delta_{\rm
  L}(\bm{k})= D(z)\delta_0(\bm{k})$, where $D(z)$ is the linear growth
factor at a redshift $z$, and $\delta_0(\bm{k})$ is the linear density
contrast at the present time $z=0$. We adopt a normalization
$D(z=0)=1$ in this paper.

The evolution of the mass density field can be evaluated
perturbatively in quasi-nonlinear regime. The SPT evaluates the
perturbative kernel $F_n$ in Eulerian space order by order
\cite{fry84,GGRW86,JB94,BCGS02}. For $n =1,2$, for example, we have
\begin{align}
  F_1(\bm{k}) &= 1,
\label{eq:1-4a}\\
  F_2(\bm{k}_1,\bm{k}_2) &= \frac{10}{7} + \left(\frac{k_1}{k_2} +
      \frac{k_1}{k_2}\right) \frac{\bm{k}_1\cdot\bm{k}_2}{k_1k_2} +
  \frac{4}{7} \left(\frac{\bm{k}_1\cdot\bm{k}_2}{k_1k_2}\right)^2.
\label{eq:1-4b}
\end{align}
Expressions of $F_3$ and $F_4$ are explicitly given in
Ref.~\cite{GGRW86}. Although above kernels $F_n$ $(n\ge 2)$ are exact
only in the Einstein--de~Sitter universe, $\varOmega_{\rm M} = 1$,
$\varOmega_\Lambda = 0$, those are good approximations in other
cosmological models with $\varOmega_{\rm M} \ne 1$, $\varOmega_\Lambda
\ne 0$ \cite{BCGS02}.

Combining Eq.~(\ref{eq:1-2}) and Eq.~(\ref{eq:1-3}), we have a formal
expansion of the form
\begin{multline}
  {\delta}_{\rm X}(\bm{k}) =
  \sum_{n=1}^\infty \frac{1}{n!}
  \int \frac{d^3k_1}{(2\pi)^3} \cdots \frac{d^3k_n}{(2\pi)^3}
  (2\pi)^3 \delta_{\rm D}^3(\bm{k}_{1\cdots n} - \bm{k}) \\
  \times K_n(\bm{k}_1,\ldots,\bm{k}_n) 
  {\delta}_{\rm L}(\bm{k}_1) \cdots {\delta}_{\rm L}(\bm{k}_n),
\label{eq:1-5}
\end{multline}
where
\begin{align}
  K_1(\bm{k}) &= b_1(\bm{k}),
  \label{eq:1-6a}\\
  K_2(\bm{k}_1,\bm{k}_2) &= b_1(\bm{k})
  F_2(\bm{k}_1,\bm{k}_2) + b_2(\bm{k}_1,\bm{k}_2)
\label{eq:1-6b}\\
  K_3(\bm{k}_1,\bm{k}_2,\bm{k}_3) &=
  b_1(\bm{k}) F_3(\bm{k}_1,\bm{k}_2,\bm{k}_3)
\nonumber\\
  & \quad + \left[
      b_2(\bm{k}_1,\bm{k}_{23})
      F_2(\bm{k}_2,\bm{k}_3)
      + \mbox{cyc.}
  \right]
\nonumber\\
  & \quad + 
      b_3(\bm{k}_1,\bm{k}_2,\bm{k}_3),
\label{eq:1-6c}
\end{align}
and so forth, where $\bm{k} = \bm{k}_{1\cdots n}$ in each expression
of $K_n$.

The $N$-point polyspectrum $P^{(N)}_{\rm X}$ of the field $\delta_{\rm
  X}$ is defined by
\begin{multline}
  \left\langle \delta_{\rm X}(\bm{k}_1) \cdots
      \delta_{\rm X}(\bm{k}_N) \right\rangle_{\rm c}
 = (2\pi)^3 \delta_{\rm D}^3(\bm{k}_{1\cdots N})
  P^{(N)}_{\rm X}(\bm{k}_1,\ldots,\bm{k}_N),
\label{eq:1-7}
\end{multline}
where $\langle\cdots\rangle_{\rm c}$ denotes the cumulant, which
corresponds to the connected part of the $N$-point expectation value.
The 2-point polyspectrum is the power spectrum $P_{\rm X} =
P^{(2)}_{\rm X}$. The 3- and 4-point polyspectra are the bispectrum
$B_{\rm X} = P^{(3)}_{\rm X}$ and the trispectrum $T_{\rm X} =
P^{(4)}_{\rm X}$, respectively. Substituting the Eq.~(\ref{eq:1-5})
into Eq.~(\ref{eq:1-7}), one can perturbatively evaluate the
polyspectra $P^{(N)}_{\rm X}$ in terms of the polyspectra
$P^{(n)}_{\rm L}$ of the linear density contrast $\delta_{\rm L}$,
which is similarly defined by
\begin{multline}
  \left\langle \delta_{\rm L}(\bm{k}_1) \cdots
      \delta_{\rm L}(\bm{k}_n) \right\rangle_{\rm c}
 = (2\pi)^3 \delta_{\rm D}^3(\bm{k}_{1\cdots n})
  P^{(n)}_{\rm L}(\bm{k}_1,\ldots,\bm{k}_n).
\label{eq:1-7-1}
\end{multline}
The linear polyspectra $P^{(n)}_{\rm L}$ are proportional to the
primordial spectra. When the initial density field is random Gaussian,
only the primordial power spectrum is present and higher-order
linear polyspectra all vanish.

\begin{figure}
\begin{center}
\includegraphics[width=18pc]{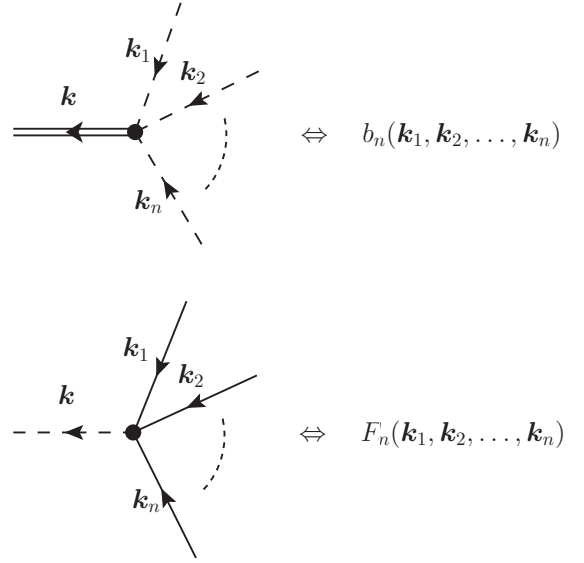}%{EPTreal.eps}
\caption{\label{fig:EPTreal} Diagrammatic rules for the Eulerian
  perturbation theory in real space. In each vertex, $\bm{k} =
  \bm{k}_1 + \cdots + \bm{k}_n$ should be satisfied. The upper
    and lower rules correspond to the expansions in Eq.~(\ref{eq:1-2})
    and (\ref{eq:1-3}), respectively. A dashed line should be
    ``internal'': one end of a dashed line should be connected
    to a vertex with double solid line, and the other end should be
    connected to a vertex with solid lines.}
\end{center}
\end{figure}

In calculating the polyspectra, diagrammatic methods are quite useful.
Fig.~\ref{fig:EPTreal} shows the diagrammatic rules for the Eulerian
perturbation theory in real space.
The first rule corresponds to each term in the expansion of
Eq.~(\ref{eq:1-2}), and the second rule corresponds to each term in
Eq.~(\ref{eq:1-3}). In both vertices, a momentum conservation $\bm{k}
= \bm{k}_1 + \cdots +\bm{k}_n$ should be satisfied, according to the
Dirac's delta function in each corresponding equation. The double
solid line, dashed line, and single solid line correspond to
$\delta_{\rm X}$, $\delta_{\rm m}$, and $\delta_{\rm L}$,
respectively. Since the variable $\delta_{\rm m}$ is expanded
according to Eq.~(\ref{eq:1-3}), dashed lines in the upper rule should
always be connected to the vertices of the lower rule, i.e., the
dashed lines are ``internal.''

\begin{figure}
\begin{center}
\includegraphics[width=18pc]{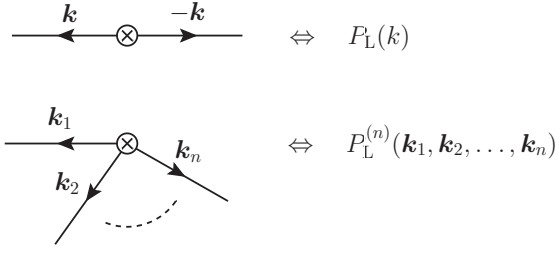}%{linspec.eps}
\caption{\label{fig:linspec} Diagrammatic rules for contributions from
  the primordial polyspectra. All the free ends of solid lines in
  Fig.~\ref{fig:EPTreal} should be connected to each other by these
  rules. When the initial density field is random Gaussian, the lower
  graph with $n\geq 3$ does not exist. In the lower graph, $\bm{k}_1 +
  \cdots + \bm{k}_n = \bm{0}$ should be satisfied. The case of $n=2$
  in the lower graph is equivalent to the upper graph.}
\end{center}
\end{figure}

To evaluate the $N$-point polyspectra of Eq.~(\ref{eq:1-7}) with
expansions of Eqs.~(\ref{eq:1-2}) and (\ref{eq:1-3}), we need
cumulants of the linear density contrast $\delta_{\rm L}$, which are
given by Eq.~(\ref{eq:1-7-1}). This procedure is diagrammatically
equivalent to applying rules in Fig.~\ref{fig:linspec}.
The open circles with cross represent the primordial polyspectra. When
the initial density field is random Gaussian, higher-order polyspectra
all vanish, $P_{\rm L}^{(n)} =0$ for $n \geq 3$, and only the upper
rule in Fig.~\ref{fig:linspec} is relevant.

For the evaluations of $N$-point polyspectra $P_{\rm X}^{(N)}$ in
Eq.~(\ref{eq:1-7}), we first consider $N$ vertices with double solid
lines. Next we consider possible ways of connecting those vertices
with dashed lines and solid lines according to the rules in
Figs.~\ref{fig:EPTreal} and \ref{fig:linspec}. The polyspectra $P_{\rm
  X}^{(N)}$ is given by the sum of terms which correspond to all the
possible diagrams, with appropriate statistical factors which are
explained at the end of this subsection.

Since the $N$-point polyspectra are defined by the connected part in
Eq.~(\ref{eq:1-7}), only connected diagrams should be taken into
account. Discarding unconnected diagrams is equivalent to taking the
connected part. When there exist internal wavevectors
$\bm{k}_1',\bm{k}_2',\ldots$, which are not uniquely determined from
external wavevectors $\bm{k}_1,\ldots,\bm{k}_N$, those internal
wavevectors should be integrated with a weight of $(2\pi)^{-3}$, i.e.,
$\int d^3k_1'/(2\pi)^3\cdot d^3k_2'/(2\pi)^3 \cdots$. The number of
internal wavevectors to be integrated is the same as the number of
loops in a given diagram.

Every terms in the perturbative expansion of the polyspectra
$P^{(N)}_{\rm X}$ of Eq.~(\ref{eq:1-7}) corresponds to the diagrams
constructed by above rules. When the hierarchical orders for the
linear polyspectra $P^{(N)}_{\rm L} \sim {\cal O}(P_{\rm L})^{N-1}$
hold, the number of loops is equal to the order of $P_{\rm L}$ in a
given diagram. In any case, the perturbative order in a given diagram
is apparent from the number and kind of crossed circles.

When the mixed polyspectra of different types of objects, such as
cross power spectra $P_{\rm mX}(k)$, $P_{\rm X_1X_2}(k)$, etc., are
need to be evaluated, we just use different vertices with
corresponding set of bias functions.

When the bias is not present, $b_1=1$, $b_2=b_3=\cdots=0$, only one
dashed line can be connected to each vertex of the lower rule in
Fig.~\ref{fig:EPTreal}. In this case one does not need to consider the
dashed line at all, and the diagrammatic rules of
Figs.~\ref{fig:EPTreal} and \ref{fig:linspec} are equivalent to the
ones which were previously introduced in Ref.~\cite{GGRW86}, in which
Gaussian initial conditions are assumed. Therefore, our diagrammatic
rules of Figs.~\ref{fig:EPTreal} and \ref{fig:linspec} are
generalization of the previous rules to the case when the Eulerian
nonlocal bias and primordial non-Gaussianity are present in general.

\begin{figure}
\begin{center}
\includegraphics[width=18pc]{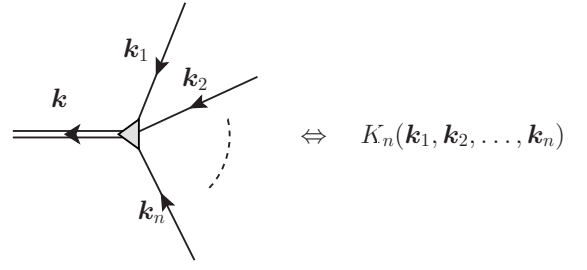}%{Kernel.eps}
\caption{\label{fig:ShrunkKernel} Shrunk vertex. The triangle
  represents all the possible tree graphs constructed by the rules in
  Fig.~\ref{fig:EPTreal} of EPT. The shrunk vertex can also be
  expressed by LPT diagrams. }
\end{center}
\end{figure}

\begin{figure}
\begin{center}
\includegraphics[width=20pc]{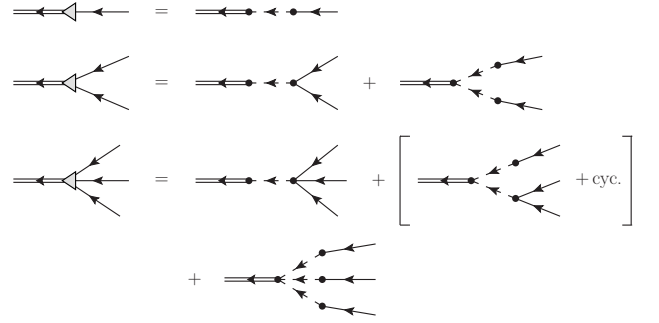}%{EPTShrinkReal.eps}
\caption{\label{fig:EPTShrinkReal} 
Shrunk vertices in Eulerian perturbation theory in real space.
}
\end{center}
\end{figure}

The diagrammatic rule for the expansion of Eq.~(\ref{eq:1-5}) is shown
in Fig.~\ref{fig:ShrunkKernel}.
The triangle vertex corresponds to shrinking the vertex in terms of
the diagrammatic rules of Fig.~\ref{fig:EPTreal}. In fact, for
$n=1,2,3$, the shrunk vertices are diagrammatically given by
Fig.~\ref{fig:EPTShrinkReal}.
These diagrammatic representations are equivalent to
Eqs.~(\ref{eq:1-6a})--(\ref{eq:1-6c}).

\begin{figure}
\begin{center}
\includegraphics[width=20pc]{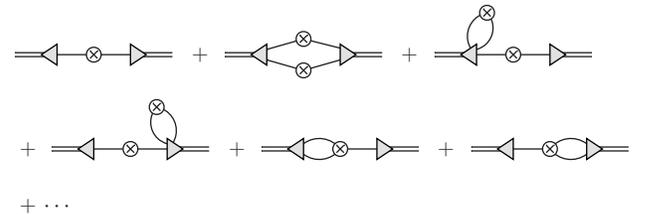}%{PspecEx.eps}
\caption{\label{fig:OneLoopPS} Diagrams up to one-loops for the power
  spectrum. }
\end{center}
\end{figure}

As an example, Fig.~\ref{fig:OneLoopPS} shows diagrams for the power
spectrum up to the one-loop order in terms of shrunk vertices.
It is a good exercise for readers to explicitly calculate the power
spectrum from Eqs.~(\ref{eq:1-5}) and (\ref{eq:1-7}), without
resorting diagrams, and compare the result with the diagrammatic
representation of Fig.~\ref{fig:OneLoopPS}. The result is given by
\begin{multline}
  P_{\rm X}(k) = [K_1(k)]^2 P_{\rm L}(k) 
\\
  + \frac12 \int \frac{d^3k'}{(2\pi)^3}
  \left[K_2(\bm{k}',\bm{k}-\bm{k}')\right]^2
      P_{\rm L}(k') P_{\rm L}(|\bm{k}-\bm{k}'|) 
\\
  + K_1(k) P_{\rm L}(k)
  \int \frac{d^3k'}{(2\pi)^3} K_3(\bm{k},\bm{k}',-\bm{k}') P_{\rm
    L}(k')
\\
  + K_1(k) \int \frac{d^3k'}{(2\pi)^3} K_2(\bm{k}',\bm{k}-\bm{k}')
    B_{\rm L}(\bm{k},\bm{k}'-\bm{k},-\bm{k}').
\label{eq:1-8}
\end{multline}

One should be careful to put a correct statistical factor in each
diagram. When there are equivalent pieces in a diagram, one should put
a statistical factor $1/n_{\rm equiv}!$, where $n_{\rm equiv}$ is the
number of that equivalent pieces. For example, the second diagram in
Fig.~\ref{fig:OneLoopPS} has a pair of equivalent lines connecting
both vertices. It has a two equivalent pieces, and the resulting
statistical factor is $1/2! = 1/2$. In another way of viewing the
statistical factor of this diagram, each vertex has a factor $1/2!$
because of the prefactor $1/n!$ in Eq.~(\ref{eq:1-5}), and there is
two ways of connecting the solid lines from the vertices, resulting in
the final statistical factor of $(1/2!)^2 \times 2 = 1/2$, which
agrees with the previous consideration of equivalent pieces. In the
third diagram, two solid lines in the loop connected to the left
vertex are equivalent pieces, and the statistical factor is $1/2! =
1/2$. In another way of viewing, left vertex has a factor $1/3!$, and
there are three ways of choosing which solid lines are associated to
the loop, resulting in the final statistical factor of $1/3! \times 3
= 1/2$, which agrees with the previous consideration of equivalent
pieces. The fourth diagram gives the same expression with the third,
due to the parity symmetry. Similarly, the fifth and sixth diagrams
have the statistical factor of $1/2! = 1/2$ in both ways of viewing.
After some experience, one can put a correct statistical factor in a
given diagram.  Two ways of counting as in the above examples are
helpful for cross-checking.

\subsection{\label{subsec:LagrangianBias}
Nonlocal bias in Lagrangian space
}

In the Lagrangian view, the dynamical evolution of cosmological
density fields is tracked by a set of trajectories of mass element,
$\bm{x}(\bm{q},t)$, where $\bm{q}$ is the initial Lagrangian
coordinates of the trajectory. A displacement field
$\bm{\varPsi}(\bm{q},t)$ is defined by
\begin{equation}
  \bm{x}(\bm{q},t) = \bm{q} + \bm{\varPsi}(\bm{q},t),
\label{eq:1-101}
\end{equation}
and is considered as a fundamental quantity in the Lagrangian view of
perturbations.

Since the initial density field is sufficiently uniform, the Eulerian
mass density field $\rho_{\rm m}(\bm{x})$ satisfies the continuity
relation,
\begin{equation}
  \rho_{\rm m}(\bm{x}) d^3x = \bar{\rho}_{\rm m} d^3q, 
\label{eq:1-102}
\end{equation}
where $\bar{\rho}_{\rm m}$ is the comoving mean density of mass. On
the other hand, the fluid elements in which observed objects reside
are not uniformly distributed in Lagrangian space. The continuity
relation is given by
\begin{equation}
  \rho_{\rm X}(\bm{x}) d^3x = {\rho}^{\rm L}_{\rm X}(\bm{q}) d^3q, 
\label{eq:1-103}
\end{equation}
where $\rho^{\rm L}_{\rm X}(\bm{q})$ is the density field of the
observed objects in Lagrangian space. Note that both $\rho_{\rm X}$
and $\rho^{\rm L}_{\rm X}$ depend on the time of observation, because
objects are identified by observers at a given time. The comoving mean
density of the objects, $\bar{\rho}_{\rm X}$ is common to both density
fields, and the Eq.~(\ref{eq:1-103}) is equivalent to the following
equation:
\begin{equation}
  1 + \delta_{\rm X}(\bm{x}) =
  \int d^3q  \left[1 + \delta^{\rm L}_{\rm X}(\bm{q})\right]
      \delta_{\rm D}^3[\bm{x}-\bm{q}-\bm{\varPsi}(\bm{q})].
\label{eq:1-104}
\end{equation}
When the density field is not biased, $\delta^{\rm L}_{\rm X}(\bm{q})
= 0$ everywhere. 

The expansion of the number density field by the linear density field
in Lagrangian space is formally given by
\begin{multline}
  {\delta}^{\rm L}_{\rm X}(\bm{k}) =
  \sum_{n=1}^\infty \frac{1}{n!}
  \int \frac{d^3k_1}{(2\pi)^3} \cdots \frac{d^3k_n}{(2\pi)^3}
  (2\pi)^3 \delta_{\rm D}^3(\bm{k}_{1\cdots n} - \bm{k}) \\
  \times b^{\rm L}_n(\bm{k}_1,\ldots,\bm{k}_n)\,
  \delta_{\rm L}(\bm{k}_1) \cdots \delta_{\rm L}(\bm{k}_n),
\label{eq:1-106}
\end{multline}
where $b^{\rm L}_n$ is the $n$-th order nonlocal bias function in
Lagrangian space.
The bias functions are essentially
infinite-dimensional Taylor coefficients, and given by functional
derivatives:
\begin{equation}
    b^{\rm L}_n(\bm{k}_1,\ldots,\bm{k}_n) = 
    (2\pi)^{3n} \int \frac{d^3k'}{(2\pi)^3}
    \left.
        \frac{\delta^n\delta_{\rm X}^{\rm L}(\bm{k}')}
        {\delta\delta_{\rm L}(\bm{k}_1)\cdots\delta\delta_{\rm
            L}(\bm{k}_n)}
    \right|_{\delta_{\rm L}=0}.
\label{eq:1-106-1}
\end{equation}
Applying the Fourier transform to the Eq.~(\ref{eq:1-104}), and
expanding the exponent of the displacement field, we have
\begin{multline}
  {\delta}_{\rm X}(\bm{k})
  = \int d^3q
  e^{-i\bm{k}\cdot\bm{q}}
  \left[1 + \delta^{\rm L}_{\rm X}(\bm{q})\right]
  e^{-i\bm{k}\cdot\bm{\varPsi}(\bm{q})} - (2\pi)^3\delta_{\rm D}^3(\bm{k})
\\
  = \sum_{n+m\geq 1}^\infty \frac{(-i)^m}{n!m!}
   \int \frac{d^3k_1}{(2\pi)^3}\cdots\frac{d^3k_n}{(2\pi)^3}
   \frac{d^3k_1'}{(2\pi)^3}\cdots\frac{d^3k_m'}{(2\pi)^3}
\\
  \times
  (2\pi)^3\delta_{\rm D}^3(\bm{k}_{1\cdots n}+\bm{k}'_{1\cdots m}-\bm{k})
   b^{\rm L}_n(\bm{k}_1,\ldots,\bm{k}_n)
\\
  \times
  \delta_{\rm L}(\bm{k}_1)\cdots\delta_{\rm L}(\bm{k}_n)
  [\bm{k}\cdot{\tilde{\bm{\varPsi}}}(\bm{k}'_1)]\cdots
  [\bm{k}\cdot{\tilde{\bm{\varPsi}}}(\bm{k}'_m)],
\label{eq:1-105}
\end{multline}
where we define $b^{\rm L}_0 \equiv 1$ above just for $n=0$, and
$\tilde{\bm{\varPsi}}$ is the Fourier transform of the displacement
field. In the following we use the displacement field both in
configuration space and Fourier space, and we notationally distinguish
between $\bm{\varPsi}$ and $\tilde{\bm{\varPsi}}$.

The displacement field $\bm{\varPsi}$ is similarly expanded
in Lagrangian space:
\begin{multline}
  \tilde{\bm{\varPsi}}(\bm{k}) =
  \sum_{n=1}^\infty \frac{i}{n!}
  \int \frac{d^3k_1}{(2\pi)^3} \cdots \frac{d^3k_n}{(2\pi)^3}
  (2\pi)^3 \delta_{\rm D}^3(\bm{k}_{1\cdots n} - \bm{k}) \\
  \times \bm{L}_n(\bm{k}_1,\ldots,\bm{k}_n)\,
  \delta_{\rm L}(\bm{k}_1) \cdots \delta_{\rm L}(\bm{k}_n).
\label{eq:1-107}
\end{multline}
The evolution of the displacement field can be perturbatively
evaluated in quasi-nonlinear regime. The LPT evaluates the
perturbative kernel $\bm{L}_n$ order by order
\cite{cat95,CT96,mat08a}. For $n=1,2$, we have
\begin{align}
  \bm{L}_1(\bm{k}) &= \frac{\bm{k}}{k^2},
  \label{eq:1-108a}\\
  \bm{L}_2(\bm{k}_1,\bm{k}_2) &=
  \frac{3}{7} \frac{\bm{k}}{k^2}
  \left[1 - \left(\frac{\bm{k}_1\cdot\bm{k}_2}{k_1k_2}\right)^2\right],
\label{eq:1-108b}
\end{align}
where $\bm{k} = \bm{k}_1 + \bm{k}_2$ for $\bm{L}_2$. As in
Eq.~(\ref{eq:1-4b}), the above kernel $\bm{L}_2$ is exact only for
Einstein--de-Sitter universe, and the expression is a good
approximation in other cosmological models. The explicit form of the
third-order kernel $\bm{L}_3$ is given in \cite{cat95}. The kernel
$\bm{L}_n$ is not proportional to $\bm{k}$ for $n\geq 3$, in general.

\begin{figure}
\begin{center}
\includegraphics[width=18pc]{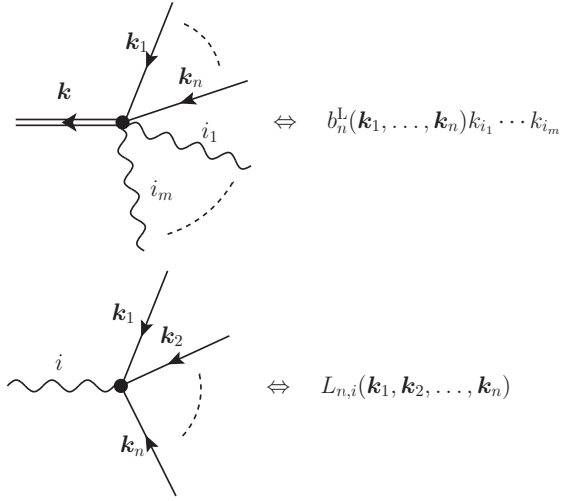}%{LPTdiag.eps}
\caption{\label{fig:LPTdiag} Diagrammatic rules for the Lagrangian
  perturbation theory in real space. The symbols $k_i$ and $L_{n,i}$
  are $i$-components of $\bm{k}$ and $\bm{L}_n$, respectively. The
  wavy lines are internal, and also carry wavevectors. The momentum
  conservation should be satisfied at each vertex. The diagrammatic
  rules in redshift space are just given by replacements $\bm{L}_n
  \rightarrow \bm{L}^{\rm s}_n$}.
\end{center}
\end{figure}

Diagrammatic rules for Eqs.~(\ref{eq:1-105}) and (\ref{eq:1-107}) are
given in Fig.~\ref{fig:LPTdiag}.
The double solid line and single solid line correspond to $\delta_{\rm
  X}$ and $\delta_{\rm L}$, respectively. The wavy line correspond to
the $n$-th order perturbation of the displacement vector
$\bm{\varPsi}$, and has an index $i$ which corresponds to a component
of the vector. The wavy lines also carry wavevectors, and
  momentum conservations should be satisfied at every vertices.

\begin{figure*}
\begin{center}
\includegraphics[width=42pc]{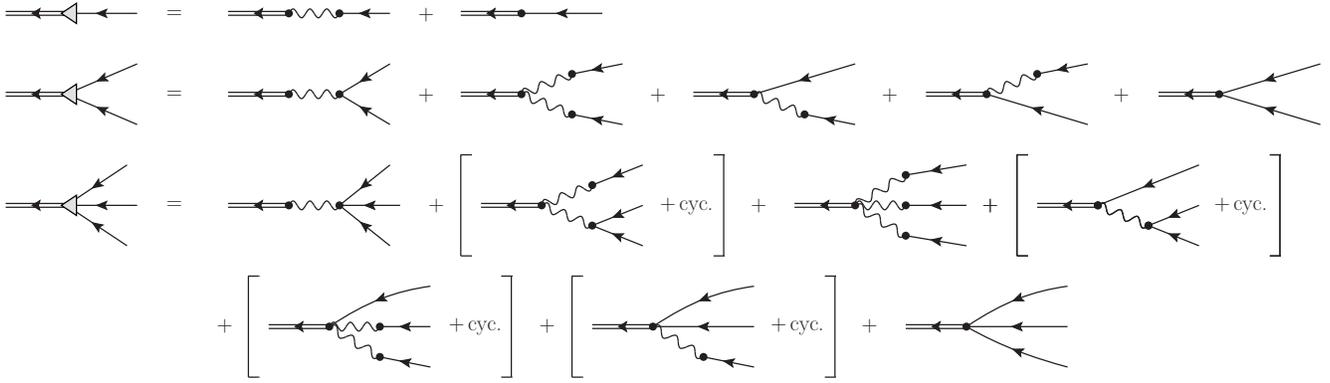}%{LPTShrink.eps}
\caption{\label{fig:LPTShrink} Shrunk vertices in terms of the
  Lagrangian perturbation theory. }
\end{center}
\end{figure*}

Combining Eqs.~(\ref{eq:1-105}) and (\ref{eq:1-107}), we have a formal
expansion which should be identical to the Eq.~(\ref{eq:1-5}) in EPT.
The first several kernels are given by
\begin{align}
  K_1(\bm{k}) &= \bm{k}\cdot\bm{L}_1(\bm{k}) + b^{\rm L}_1(\bm{k}),
  \label{eq:1-109a}\\
  K_2(\bm{k}_1,\bm{k}_2) &=
  \bm{k}\cdot\bm{L}_2(\bm{k}_1,\bm{k}_2) +
  [\bm{k}\cdot\bm{L}_1(\bm{k}_1)] [\bm{k}\cdot\bm{L}_1(\bm{k}_2)] 
\nonumber\\
  & \quad +
      b^{\rm L}_1(\bm{k}_1)[\bm{k}\cdot\bm{L}_1(\bm{k}_2)]
      + b^{\rm L}_1(\bm{k}_2)[\bm{k}\cdot\bm{L}_1(\bm{k}_1)]
\nonumber\\
  & \quad + b^{\rm L}_2(\bm{k}_1,\bm{k}_2),
\label{eq:1-109b}\\
  K_3(\bm{k}_1,\bm{k}_2,\bm{k}_3) &=
  \bm{k}\cdot\bm{L}_3(\bm{k}_1,\bm{k}_2,\bm{k}_3)
\nonumber\\
  & \quad + \left\{
      [\bm{k}\cdot\bm{L}_1(\bm{k}_1)]
      [\bm{k}\cdot\bm{L}_2(\bm{k}_2,\bm{k}_3)] + \mbox{cyc.}
      \right\}
\nonumber\\
  & \quad + 
      [\bm{k}\cdot\bm{L}_1(\bm{k}_1)]
      [\bm{k}\cdot\bm{L}_1(\bm{k}_2)]
      [\bm{k}\cdot\bm{L}_1(\bm{k}_3)]
\nonumber\\
  & \quad + \left\{
      b^{\rm L}_1(\bm{k}_1)[\bm{k}\cdot\bm{L}_2(\bm{k}_2,\bm{k}_3)]
      + \mbox{cyc.}
      \right\}
\nonumber\\
  & \quad + \left\{
      b^{\rm L}_1(\bm{k}_1)[\bm{k}\cdot\bm{L}_1(\bm{k}_2)]
      [\bm{k}\cdot\bm{L}_1(\bm{k}_3)]
      + \mbox{cyc.}
      \right\}
\nonumber\\
  & \quad + \left\{
      b^{\rm L}_2(\bm{k}_1,\bm{k}_2)[\bm{k}\cdot\bm{L}_1(\bm{k}_3)]
      + \mbox{cyc.}
      \right\}
\nonumber\\
  & \quad + b^{\rm L}_3(\bm{k}_1,\bm{k}_2,\bm{k}_3)
\label{eq:1-109c}
\end{align}
and so forth, where $\bm{k} = \bm{k}_{1\cdots n}$ for $K_n$. Those
equations are diagrammatically represented in
Fig.~\ref{fig:LPTShrink}.

When the bias is not present, $b_1 = 1$, $b^{\rm L}_1 = 0$, and
$b_n=b^{\rm L}_n = 0$ for $n\geq 2$. In this case, the equivalence of
Eqs.~(\ref{eq:1-6a})--(\ref{eq:1-6c}) and
Eqs.~(\ref{eq:1-109a})--(\ref{eq:1-109c}) indicates
\begin{align}
  F_1(\bm{k}) &= \bm{k}\cdot\bm{L}_1(\bm{k}),
  \label{eq:1-110a}\\
  F_2(\bm{k}_1,\bm{k}_2) &=
  \bm{k}\cdot\bm{L}_2(\bm{k}_1,\bm{k}_2) +
  [\bm{k}\cdot\bm{L}_1(\bm{k}_1)] [\bm{k}\cdot\bm{L}_1(\bm{k}_2)],
\label{eq:1-110b}\\
  F_3(\bm{k}_1,\bm{k}_2,\bm{k}_3) &=
  \bm{k}\cdot\bm{L}_3(\bm{k}_1,\bm{k}_2,\bm{k}_3)
\nonumber\\
  & \quad + \left\{
      [\bm{k}\cdot\bm{L}_1(\bm{k}_1)]
      [\bm{k}\cdot\bm{L}_2(\bm{k}_2,\bm{k}_3)] + \mbox{cyc.}
      \right\}
\nonumber\\
  & \quad + 
      [\bm{k}\cdot\bm{L}_1(\bm{k}_1)]
      [\bm{k}\cdot\bm{L}_1(\bm{k}_2)]
      [\bm{k}\cdot\bm{L}_1(\bm{k}_3)],
\label{eq:1-110c}
\end{align}
and so forth, where $\bm{k} = \bm{k}_{1\cdots n}$ for $F_n$. The above
equations give the relations of kernels in Lagrangian and Eulerian
perturbation theories for dark matter evolutions. Therefore, those
equations are identities. It is easily seen that
Eqs.~(\ref{eq:1-110a}), (\ref{eq:1-110b}) explicitly hold for
Eqs.~(\ref{eq:1-4a}), (\ref{eq:1-4b}), (\ref{eq:1-108a}),
(\ref{eq:1-108b}).

\subsection{\label{subsec:ELBrel}
The relation between Eulerian and Lagrangian bias functions
}

In the presence of bias, the equivalence of
Eqs.~(\ref{eq:1-6a})--(\ref{eq:1-6c}) and
Eqs.~(\ref{eq:1-109a})--(\ref{eq:1-109c}) indicates the relation
between the Eulerian and Lagrangian bias. First several relations are
given by
\begin{align}
  b_1(\bm{k}) &= b^{\rm L}_1(\bm{k})+1,
\label{eq:1-111a}\\
  b_2(\bm{k}_1,\bm{k}_2) &= b^{\rm L}_2(\bm{k}_1,\bm{k}_2)
  - F_2(\bm{k}_1,\bm{k}_2) b^{\rm L}_1(\bm{k}_{12})
\nonumber\\
  & \quad + \left\{
      \left[\bm{k}\cdot\bm{L}_1(\bm{k}_1)\right] b^{\rm L}_1(\bm{k}_2) 
      + (\bm{k}_1 \leftrightarrow \bm{k}_2)
  \right\},
\label{eq:1-111b}\\
  b_3(\bm{k}_1,\bm{k}_2,\bm{k}_3) &=
  b^{\rm L}_3(\bm{k}_1,\bm{k}_2,\bm{k}_3)
\nonumber\\
  & \quad
  - \left[
      F_2(\bm{k}_1,\bm{k}_2) b_2^{\rm L}(\bm{k}_{12},\bm{k}_3) +
      \mbox{cyc.} 
    \right]
\nonumber\\
  & \quad
  + \left[
      \bm{k}\cdot\bm{L}_1(\bm{k}_3) b_2^{\rm L}(\bm{k}_1,\bm{k}_2) +
      \mbox{cyc.} 
    \right],
\nonumber\\
  & \quad +
  \bigl[
      F_2(\bm{k}_1,\bm{k}_2) F_2(\bm{k}_{12},\bm{k}_3) + \mbox{cyc.}
\nonumber\\
  & \hspace{7pc}
      - F_3(\bm{k}_1,\bm{k}_2,\bm{k}_3)
  \bigr]b_1^{\rm L}(\bm{k})
\nonumber\\
  & \quad
  - \left[ \bm{k}\cdot\bm{L}_1(\bm{k}_3) F_2(\bm{k}_1,\bm{k}_2)
      b_1^{\rm L}(\bm{k}_{12}) \ + \mbox{cyc.}\right]
\nonumber\\
  & \quad
  + \Bigl\{
      \bigl[\bm{k}\cdot\bm{L}_2(\bm{k}_1,\bm{k}_2)
      + \bm{k}\cdot\bm{L}_1(\bm{k}_1) \bm{k}\cdot\bm{L}_1(\bm{k}_2)
\nonumber\\
  & \qquad\quad
      - \bm{k}\cdot\bm{L}_1(\bm{k}_{12}) F_2(\bm{k}_1,\bm{k}_2)
  \bigr] b_1^{\rm L}(\bm{k}_3) + \mbox{cyc.}\Bigr\},
\label{eq:1-111c}
\end{align}
and so forth, where $\bm{k} = \bm{k}_{1\cdots n}$ for $b_n$.

% The above
% equations are straightforwardly inverted to give
% \begin{align}
%   b^{\rm L}_1(\bm{k}) &= b_1(\bm{k}) - 1
% \label{eq:1-112a}\\
%   b^{\rm L}_2(\bm{k}_1,\bm{k}_2) &= b_2(\bm{k}_1,\bm{k}_2)
%   + F_2(\bm{k}_1,\bm{k}_2) b_1(\bm{k}_{12})
% \nonumber\\
%   &\quad
%   - \left(1 + \frac{\bm{k}_1\cdot\bm{k}_2}{k_1^2}\right) b_1(\bm{k}_2) 
%   - \left(1 + \frac{\bm{k}_1\cdot\bm{k}_2}{k_2^2}\right) b_1(\bm{k}_1)
% \nonumber\\
%   &\quad
%   + \frac47
%   \left[1 - \left(\frac{\bm{k}_1\cdot\bm{k}_2}{k_1k_2}\right)^2\right],
% \label{eq:1-112b}
% \end{align}
% and so forth, where explicit forms of kernel,
% Eqs.~(\ref{eq:1-4a}), (\ref{eq:1-4b}), (\ref{eq:1-108a}),
% (\ref{eq:1-108b}) are used.

An immediate consequence of the above formulas is that the biasing
cannot be local simultaneously both in Eulerian and in Lagrangian
space, since the bias parameters are all constants in local bias
models. For example, when the Lagrangian bias is local and parameters
$b_1^{\rm L}, b_2^{\rm L}, \ldots$ are scale-independent constants,
the higher-order Eulerian parameters $b_2, b_3, \ldots$ are inevitably
scale-dependent according to
Eqs.~(\ref{eq:1-111a})--(\ref{eq:1-111c}). The reason for the
incompatibility of local biases is that nonlinear evolutions are
nonlocal process in general, as we already described in Introduction.

In spite of that, local relations between the Lagrangian halo bias and
the local Eulerian bias parameters are known in the halo model
\cite{CS02}. These relations are derived by applying the spherical
collapse model \cite{MW96,MJW97}, in which the density evolutions are
local process, and therefore the local biases are compatible. This
compatibility does not hold in generally non-spherical collapse
\cite{mat08b}.

\subsection{\label{subsec:ComSph} Comments on the bias relations in
  the spherical collapse model }

The purpose of this subsection is to find a relation between the bias
parameters of general perturbation theory and that of spherical
collapse model. As mentioned above, the spherical collapse model is
used in the halo approach to find a local relation between the
(Lagrangian) halo bias parameters and Eulerian bias parameters.

Following the same manner of Ref.~\cite{MJW97}, one can derive general
relations between local bias parameters in Lagrangian space and in
Eulerian space. Such relations are derived in
Appendix~\ref{app:SphericalModel}. Up to the third order, we have
\begin{align}
  b_1 &= b^{\rm L}_1+1,
\label{eq:1-113a}\\
  b_2 &= b^{\rm L}_2 + \frac{8}{21} b^{\rm L}_1,
\label{eq:1-113b}\\
  b_3 &= b^{\rm L}_3  - \frac{13}{7} b^{\rm L}_2 - \frac{796}{1323}
  b^{\rm L}_1,
\label{eq:1-113c}
\end{align}
where both the Eulerian bias parameters $\{b_n\}$ and the Lagrangian
bias parameters $\{b^{\rm L}_n\}$ are local and scale-independent at the
same time.

Remarkably, the local bias relations in
Eqs.~(\ref{eq:1-113a})--(\ref{eq:1-113c}) for the spherical model can
be derived from Eqs.~(\ref{eq:1-111a})--(\ref{eq:1-111b}) when the
Lagrangian bias is local and Eulerian bias functions are averaged over
directions of wavevectors. Each bias function $b^{\rm
  L}_n(\bm{k}_1,\ldots,\bm{k}_n)$ is replaced by a constant $b^{\rm
  L}_n$ in the local Lagrangian bias. Angular averages of
Eqs.~(\ref{eq:1-111a})--(\ref{eq:1-111c}) in this case are
straightforwardly calculated by using the explicit forms of $F_2$,
$\bm{L}_1$ and $\bm{L}_2$ in Eqs.~(\ref{eq:1-4b}), (\ref{eq:1-108a})
and (\ref{eq:1-108b}). The angular average of $F_3$ is given by
$\langle F_3 \rangle = 682/189$ \cite{ber92,BCGS02}. As a result, the
angular averages of Eqs.~(\ref{eq:1-111a})--(\ref{eq:1-111c}) exactly
reduce to the right-hand sides of
Eqs.~(\ref{eq:1-113a})--(\ref{eq:1-113c}).

Therefore, the scale- and angular-dependences of bias are neglected in
the bias relations of the spherical collapse model, which are widely
used in the halo approach. It is only when those dependences of bias
are not important that the local bias relations of
Eqs.~(\ref{eq:1-113a})--(\ref{eq:1-113b}) are useful. In general
perturbations without spherical symmetry, one should use the nonlocal
relations of bias in Eqs.~(\ref{eq:1-111a})--(\ref{eq:1-111c}).

\subsection{\label{subsec:ComStoc} Comments on the stochastic bias}

In the framework of local bias models, the deterministic property of
bias is not viable in reality. The number density of observed objects
is not solely determined by the local density of mass. The stochastic
biasing scheme \cite{DL99} is a phenomenological model to treat the
biasing as a nondeterministic process, in the framework of local bias
models.

One of the characteristic parameters of stochasticity is the
correlation coefficient, defined by
\begin{equation}
    r(R) = \frac{\left\langle\delta_{\rm m}(R)\delta_{\rm X}(R)\right\rangle}
    {\sigma_{\rm m}(R)\sigma_{\rm X}(R)},
\label{eq:1-201}
\end{equation}
where $\sigma_{\rm m}(R) = \langle\delta_{\rm m}^2(R)\rangle^{1/2}$,
$\sigma_{\rm X}(R) = \langle\delta_{\rm X}^2(R)\rangle^{1/2}$ are
square roots of the variances of density contrasts $\delta_{\rm
  m}(R)$, $\delta_{\rm X}(R)$ which are smoothed by a radius $R$ at a
point in Eulerian space. If the deterministic relation $\delta_{\rm
  X}(R) = b(R) \delta_{\rm m}(R)$ exactly holds, the stochasticity
parameter $r(R)$ is identically unity. Deviations of the stochasticity
parameter from unity characterize how the stochasticity is important.
It is also common to define the correlation coefficient in Fourier
space,
\begin{equation}
    r(k) = \frac{P_{\rm mX}(k)}{\sqrt{P_{\rm m}(k) P_{\rm X}(k)}},
\label{eq:1-202}
\end{equation}
where $P_{\rm mX}(k)$ is the cross power spectrum of mass and objects,
$P_{\rm m}(k)$ and $P_{\rm X}(k)$ are the power spectra of mass and
objects, respectively.

This approach is purely phenomenological in the sense that the
stochasticity itself does not correspond to any fundamental physics.
Instead, the stochasticity represents our ignorance on the formation
process of observed objects. Dynamical evolutions of density field and
the formation process of observed objects are deterministic at the
fundamental level. If we would know all the detailed physics of the
formation process, any stochastic character should not appear when we
properly describe precise dependences of the number density of objects
on physical quantities, not only a local density of mass.

For the above reason, the parameters of stochastic bias should be
derived from the nonlocal bias functions,
$b_n(\bm{k}_1,\ldots,\bm{k}_n)$. At linear order, the correlation
coefficient in Fourier space $r(k)$ is identically unity, since the
bias is multiplicative, $\delta_{\rm X}(\bm{k}) = b_1(k)\delta_{\rm
  m}(\bm{k})$, which is a consequence of the translational invariance.
Even in this case, the correlation coefficient in configuration space,
Eq.~(\ref{eq:1-201}), is less than unity in general when the linear
bias parameter $b_1(k)$ is scale-dependent \cite{mat99,DS10}.

At nonlinear orders, the correlation coefficient even in Fourier space
becomes less than unity. It is straightforward to calculate the
Eq.~(\ref{eq:1-202}) in our framework of nonlocal biasing. The
lowest-order contribution to $1-r(k)$ is given by one-loop diagrams.
The relevant diagrams are similar to Fig.~\ref{fig:OneLoopPS}, and the
final result simply reduces to
\begin{multline}
    1 - r(k) = \frac{1}{4 [b_1(k)]^2 P_{\rm L}(k)}
\\ \times
    \int\frac{d^3k'}{(2\pi)^3}
    \left[b_2(\bm{k}',\bm{k}-\bm{k}')\right]^2
    P_{\rm L}(k') P_{\rm L}(|\bm{k}-\bm{k}'|).
\label{eq:1-203}
\end{multline}
This result is valid even when the initial density field is
non-Gaussian. The bispectrum contributions to one-loop power spectra
cancel out in the combination of Eq.~(\ref{eq:1-202}).

In a case of local Lagrangian bias, including the halo bias, the
Lagrangian functions $b^{\rm L}_n$ are constants and the second-order
Eulerian bias function of Eq.~(\ref{eq:1-111b}) reduces to
\begin{equation}
  b_2(\bm{k}_1,\bm{k}_2) = b^{\rm L}_2 + \frac47
  \left[1-\left(\frac{\bm{k}_1\cdot\bm{k}_2}{k_1k_2}\right)^2 \right]
  b^{\rm L}_1.
\label{eq:1-204}
\end{equation}
In the large-scale limit, $k \rightarrow 0$, we have
$b_2(\bm{k}',\bm{k}-\bm{k}') \rightarrow b^{\rm L}_2$, and
from Eq.~(\ref{eq:1-202}),
\begin{equation}
  1 - r(k) \rightarrow 
  \frac{(b^{\rm L}_2)^2}{4 [b_1(k)]^2 P_{\rm L}(k)}
    \int\frac{d^3k'}{(2\pi)^3}
    \left[P_{\rm L}(k')\right]^2.
\label{eq:1-205}
\end{equation}
It is interesting to note that the stochasticity emerges even in the
large-scale limit when the second-order Lagrangian bias parameter
$b^{\rm L}_2$ is nonzero. One cannot find such kind of properties in
an original approach of the stochastic biasing, since the correlation
coefficient is just a free parameter in the latter. The halo model
actually predicts the nonzero value of the second-order bias parameter
$b^{\rm L}_2$ (see Sec.~\ref{subsec:HaloBias} below).

To summarize this subsection, the stochastic bias is a phenomenology
which is conveniently introduced in the context of local bias models,
and is not needed in nonlocal bias models. Stochastic properties of
the local bias are derived from the deterministic nonlocal bias.

\section{\label{sec:PTred}
Perturbation theory in redshift space with nonlocal bias
}

\subsection{\label{subsec:PTredEuler}
Nonlocal bias in Eulerian space and redshift-space distortions
}

The comoving position $\bm{x}$ in real space and $\bm{s}$ in redshift
space are related by \cite{mat00}
\begin{equation}
  \bm{s} = \bm{x} + \frac{v_z(\bm{x})}{aH} \hat{\bm{z}}.
\label{eq:2-21}
\end{equation}
in the plane-parallel limit of the distant-observer approximation,
where $\hat{\bm{z}}$ is the unit vector along the line of sight, and
$v_z$ is the velocity component along $\hat{\bm{z}}$. The number density
field in real space $\rho_{\rm X}(\bm{x})$ and that in redshift space
$\rho^{\rm s}_{\rm X}(\bm{s})$ are related by a continuity relation:
\begin{equation}
  \rho^{\rm s}_{\rm X}(\bm{s}) d^3s = \rho_{\rm X}(\bm{x}) d^3x,
\label{eq:2-22}
\end{equation}
Therefore, the density contrast in redshift space is given by
\begin{equation}
  \delta^{\rm s}_{\rm X}(\bm{s})
  = \left[1 + \delta_{\rm X}(\bm{x})\right]J^{-1} - 1,
\label{eq:2-23}
\end{equation}
where $J = \partial(\bm{s})/\partial(\bm{x})$ is the Jacobian of the
mapping from real space to redshift space. One can easily calculate
the Fourier transform of Eq.~(\ref{eq:2-23}):
\begin{equation}
  {\delta}^{\rm s}_{\rm X}(\bm{k})
  = \int d^3x\, e^{-i\bm{k}\cdot\bm{x}}
  \left[1 + \delta_{\rm X}(\bm{x})\right] e^{ik_z u_z(\bm{x})},
\label{eq:2-24}
\end{equation}
where $u_z = - v_z/aH$ and $\bm{k}\ne \bm{0}$ is assumed. A different
expression of the above formula is seen in the Eq.~(4) of
Ref.~\cite{SCF99}, and it can be shown by partial integration that the
two expressions are actually equivalent. The above relation is
applicable even in the fully nonlinear regime.

In Fourier space,
\begin{equation}
  {u}_z(\bm{k}) = -i k_z {\theta}(\bm{k})/k^2,
\label{eq:2-25}
\end{equation}
where $\theta(\bm{k})$ is the Fourier transform of the normalized
velocity convergence, $\theta = -\bm{\nabla}\cdot\bm{v}/aH$.
Expanding the peculiar velocity field in Eq.~(\ref{eq:2-24}), we have
\begin{multline}
  {\delta}^{\rm s}_{\rm X}(\bm{k})
  = \sum_{n+m\geq 1}^\infty \frac{(\mu k)^n}{n!m!}
   \int \frac{d^3k_1}{(2\pi)^3}\cdots\frac{d^3k_n}{(2\pi)^3}
   \frac{d^3k_1'}{(2\pi)^3}\cdots\frac{d^3k_m'}{(2\pi)^3}
\\
  \times
  (2\pi)^3\delta_{\rm D}^3(\bm{k}_{1\cdots n}+\bm{k}'_{1\cdots m}-\bm{k})
  \frac{\mu_1\cdots\mu_n}{k_1\cdots k_n}
   b_m(\bm{k}_1',\ldots,\bm{k}_m')
\\
  \times
   \theta(\bm{k}_1)\cdots\theta(\bm{k}_n)
  \delta_{\rm m}(\bm{k}_1')\cdots\delta_{\rm m}(\bm{k}_m'),
\label{eq:2-26}
\end{multline}
where we define $b_0 \equiv 1$ above just for $m=0$, and $\mu \equiv
\bm{k}\cdot\hat{\bm{z}}/k$, $\mu_i \equiv
\bm{k}_i\cdot\hat{\bm{z}}/k_i$ are direction cosines of wavevectors.
It can be shown that this equation is equivalent to the Eq.~(5) of
Ref.~\cite{SCF99}, although the apparent expressions are somewhat
different.

In the SPT, only growing-mode solutions in each order are retained,
and the peculiar velocity field is consistently assumed to be
irrotational. Thus, the velocity field is fully characterized by the
velocity divergence \cite{BCGS02}, which is expanded by a linear
density contrast as
\begin{multline}
  \theta(\bm{k}) =
  \sum_{n=1}^\infty \frac{f}{n!}
  \int \frac{d^3k_1}{(2\pi)^3} \cdots \frac{d^3k_n}{(2\pi)^3}
  (2\pi)^3 \delta_{\rm D}^3(\bm{k}_{1\cdots n} - \bm{k}) \\
  \times G_n(\bm{k}_1,\ldots,\bm{k}_n) 
  {\delta}_{\rm L}(\bm{k}_1) \cdots {\delta}_{\rm L}(\bm{k}_n),
\label{eq:2-27}
\end{multline}
where $f = d\ln D/d\ln a = \dot{D}/HD$ is the linear growth rate which
corresponds to the logarithmic derivative of the linear growth factor.
The perturbative kernels $G_n$ are given by SPT
\cite{fry84,JB94,GGRW86,BCGS02}. For $n=1,2$, we have
\begin{align}
  G_1(\bm{k}) &= 1,
  \label{eq:2-28a}\\
  G_2(\bm{k}_1,\bm{k}_2) &= \frac{6}{7} + \left(\frac{k_1}{k_2} +
      \frac{k_1}{k_2}\right) \frac{\bm{k}_1\cdot\bm{k}_2}{k_1k_2} +
  \frac{8}{7} \left(\frac{\bm{k}_1\cdot\bm{k}_2}{k_1k_2}\right)^2,
\label{eq:2-28b}
\end{align}
where the expression $G_2$ is exact only in Einstein--de-Sitter
universe, and only weakly depends on time in general cosmology.

\begin{figure}
\begin{center}
\includegraphics[width=20pc]{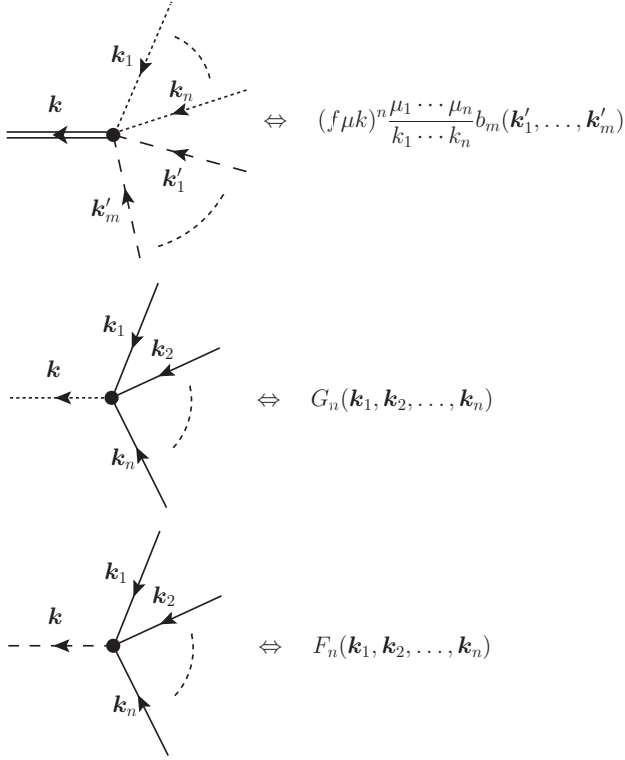}%{EPTred.eps}
\caption{\label{fig:EPTred} Diagrammatic rules for the Eulerian
  perturbation theory in redshift space. In the first rule, $\bm{k} =
  \bm{k}_1 + \cdots \bm{k}_n + \bm{k}'_1 + \cdots \bm{k}'_m$ should be
  satisfied, and $\bm{k} = \bm{k}_1 + \cdots \bm{k}_n$ should be
  satisfied in the last two rules. The dashed lines and dotted lines
  are internal.}
\end{center}
\end{figure}

Fig.~\ref{fig:EPTred} shows the diagrammatic rules for the Eulerian
perturbation theory in redshift space.
The first rule corresponds to the expansion of Eq.~(\ref{eq:2-26}),
and the second rule corresponds to Eq.~(\ref{eq:2-27}). The rules are
used in a similar way of those in real space. The third rule is common
to the rule in real space, and corresponds to Eq.~(\ref{eq:1-3}). The
momentum conservation should satisfied in each vertex. The meanings of
double solid line, dashed line and single solid line are the same as
in real space. The dotted line corresponds to the velocity convergence
$\theta$.

\begin{figure}
\begin{center}
\includegraphics[width=20pc]{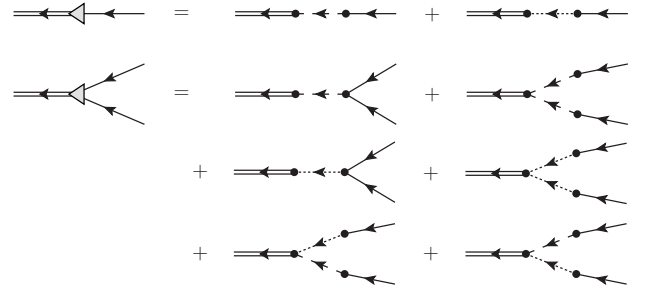}%{EPTShrinkRed.eps}
\caption{\label{fig:EPTShrinkRed} 
Shrunk vertices in Eulerian perturbation theory in redshift space.
}
\end{center}
\end{figure}

Substituting the perturbative expansions of Eqs.~(\ref{eq:1-3}) and
(\ref{eq:2-27}) into Eq.~(\ref{eq:2-26}), we have a formal series of
the biased field in redshift space:
\begin{multline}
   {\delta}^{\rm s}_{\rm X}(\bm{k}) =
    \sum_{n=1}^\infty
    \int\frac{d^3k_1}{(2\pi)^3}\cdots\frac{d^3k_n}{(2\pi)^3}
    (2\pi)^3\delta^3\left(\bm{k}_{1\cdots n} - \bm{k}\right)
\\
    \times S_n(\bm{k}_1,\ldots,\bm{k}_n)
    \delta_{\rm L}(\bm{k}_1) \cdots \delta_{\rm L}(\bm{k}_n),
\label{eq:2-29}
\end{multline}
where $S_n$'s are derived kernels. For $n=1,2$, we have
\begin{align}
    S_1(\bm{k}) &= b_1(\bm{k}) + f \mu^2,
\label{eq:2-30a}\\
    S_2(\bm{k}_1,\bm{k}_2)
    &= b_1(\bm{k}_{12}) F_2(\bm{k}_1,\bm{k}_2)
    + b_2(\bm{k}_1,\bm{k}_2)
\nonumber\\
    &\quad
    + f \mu^2 G_2(\bm{k}_1,\bm{k}_2)
    + (f \mu k)^2 \frac{\mu_1 \mu_2}{k_1 k_2}
\nonumber\\
    &\quad
    + f \mu k
    \left[\frac{\mu_1}{k_1}b_1(\bm{k}_2)
        + \frac{\mu_2}{k_2}b_1(\bm{k}_1) \right].
\label{eq:2-30b}
\end{align}
In the case of local biasing where $b_1$ and $b_2$ are constants, these
kernels are equivalent to the Eqs.~(11) and (12) in Ref.~\cite{SCF99}.
Those equations are diagrammatically represented in
Fig.~\ref{fig:EPTShrinkRed}.

\subsection{\label{subsec:PTredLagr}
Nonlocal bias in Lagrangian space and redshift-space distortions
}

Redshift-space distortions are naturally derived in the Lagrangian
picture. The velocity of a mass element with Lagrangian coordinates
$\bm{q}$ is given by a simple derivative of Eq.~(\ref{eq:1-101}):
\begin{equation}
  \bm{v}(\bm{q}) = a\dot{\bm{x}}(\bm{q}) = a\dot{\bm{\varPsi}}(\bm{q}),
\label{eq:2-50}
\end{equation}
where the dot represents a partial derivative by the proper time $t$.
Note that we suppress the argument of the time $t$ in our variables.
From Eq.~(\ref{eq:2-21}), the displacement field in redshift space is
simply given by
\begin{equation}
  \bm{\varPsi}^{\rm s} = 
  \bm{\varPsi} + \frac{\hat{\bm{z}}\cdot\dot{\bm{\varPsi}}}{H}
  \hat{\bm{z}}.
\label{eq:2-51}
\end{equation}
The redshift-space distortions are exactly linear mappings of the
displacement field, even in the nonlinear regime \cite{mat08a}. A
similar equation has been applied to the analysis of the Zel'dovich
approximation \cite{TH96}.

In our approximation that the perturbative kernels $\bm{L}_n$ are
independent on time, we have $\bm{\varPsi}^{(n)} \propto D^n$, where
$\bm{\varPsi}^{(n)}$ is the configuration-space counterpart of $n$-th
order term in Eq.(\ref{eq:1-107}). Therefore, the time derivative of
the displacement field is simply given by
\begin{equation}
  \dot{\bm{\varPsi}}^{(n)} = n H f \bm{\varPsi}^{(n)},
\label{eq:2-52}
\end{equation}
and the displacement field of each order in redshift space is
related to the real-space displacement via
\begin{equation}
 \bm{\varPsi}^{{\rm s}(n)} = 
  \bm{\varPsi}^{(n)}
  + n f (\hat{\bm{z}}\cdot\bm{\varPsi}^{(n)})\, \hat{\bm{z}},
\label{eq:2-53}
\end{equation}
which is just a linear mapping of the displacement field in each
order. This linear transformation is characterized by a redshift-space
distortion tensor $R^{(n)}_{ij}$ for each $n$ \cite{mat08a}, which is
defined by
\begin{equation}
  R^{(n)}_{ij} = \delta_{ij} + n f \hat{z}_i \hat{z}_j.
\label{eq:2-54}
\end{equation}
The Eq.~(\ref{eq:2-53}) reduces to $\varPsi^{{\rm s}(n)}_i =
R^{(n)}_{ij}\varPsi^{(n)}_j$, or in a vector notation,
$\bm{\varPsi}^{{\rm s}(n)} = R^{(n)}\bm{\varPsi}^{(n)}$.

As a result, each perturbative kernel of the redshift-space
displacement is given by a linear transformation from the real-space
kernel,
\begin{equation}
  \bm{L}^{\rm s}_n = R^{(n)} \bm{L}_n.
\label{eq:2-55}
\end{equation}
For $n=1,2$, we have
\begin{align}
  \bm{L}^{\rm s}_1(\bm{k}) &= \frac{\bm{k}+f\mu k \hat{\bm{z}}}{k^2},
  \label{eq:2-56a}\\
  \bm{L}^{\rm s}_2(\bm{k}_1,\bm{k}_2) &=
  \frac{3}{7}\,
  \frac{\bm{k} + 2 f \mu k \hat{\bm{z}}}{k^2}
 \left[1 - \left(\frac{\bm{k}_1\cdot\bm{k}_2}{k_1k_2}\right)^2\right],
\label{eq:2-56b}
\end{align}
where $\bm{k} = \bm{k}_1 + \bm{k}_2$ for $\bm{L}^{\rm s}_2$. All the
formalism of Sec.~\ref{subsec:LagrangianBias} in real space applies with
substitutions of $\bm{L}_n \rightarrow \bm{L}^{\rm s}_n$ and $K_n
\rightarrow S_n$ in redshift space. Corresponding to
Eqs.~(\ref{eq:1-109a}) and (\ref{eq:1-109b}), we have
\begin{align}
  S_1(\bm{k}) &= \bm{k}\cdot\bm{L}^{\rm s}_1(\bm{k}) + b^{\rm L}_1(\bm{k}),
  \label{eq:2-57a}\\
  S_2(\bm{k}_1,\bm{k}_2) &=
  \bm{k}\cdot\bm{L}^{\rm s}_2(\bm{k}_1,\bm{k}_2) +
  [\bm{k}\cdot\bm{L}^{\rm s}_1(\bm{k}_1)] [\bm{k}\cdot\bm{L}^{\rm s}_1(\bm{k}_2)] 
\nonumber\\
  & \quad +
      b^{\rm L}_1(\bm{k}_1)[\bm{k}\cdot\bm{L}^{\rm s}_1(\bm{k}_2)]
      + b^{\rm L}_1(\bm{k}_2)[\bm{k}\cdot\bm{L}^{\rm s}_1(\bm{k}_1)]
\nonumber\\
  & \quad + b^{\rm L}_2(\bm{k}_1,\bm{k}_2),
\label{eq:2-57b}
\end{align}
and so forth, where $\bm{k} = \bm{k}_{1\cdots n}$ for $S_n$.
Substituting Eqs.~(\ref{eq:2-56a}), (\ref{eq:2-56b}), and
Eqs.~(\ref{eq:1-111a}), (\ref{eq:1-111b}), the above equations are
equivalent to Eqs.~(\ref{eq:2-30a}), (\ref{eq:2-30b}). Thus, the
relation of the nonlocal biases of Eqs.~(\ref{eq:1-111a}),
(\ref{eq:1-111b}), which are derived in real space, also consistently
applies in redshift space. Such consistency should hold for any
higher-order kernels.

\section{\label{sec:VResum} Vertex Resummations
}

The formalism presented so far is a natural extension of the SPT,
simultaneously including the nonlocal bias, redshift-space distortions
and primordial non-Gaussianity. In recent years, various methods
beyond the SPT are developed as mentioned in Introduction. In the RPT
\cite{CS06a,CS06b}, the ``propagator'' plays an important role. The
concept of the propagator is extended to the ``multi-point
propagator'' \cite{BCS08,BCS10}, in which the original propagator is
identified as the one-point propagator.

In this section, we show that the multi-point propagators can be
obtained by resumming the external vertices in our formalism. Most of
the resummation methods known so far are only applied to the dark
matter clustering in real space. Exceptions are the Lagrangian
resummation method \cite{mat08a,mat08b}, in which local Lagrangian
bias and redshift-space distortions are included, and the
time-renormalization-group method \cite{eli10}), in which the halo
bias is included. Therefore, the identification of the multi-point
propagators in our formalism is an important step toward including
all the effects of nonlocal biasing, redshift-space distortions and
primordial non-Gaussianity into the resummation methods.

\subsection{\label{subsec:MProp} Multi-point propagators}

We first consider how the multi-point propagator \cite{BCS08} is
related to our formalism. We only consider one-component propagators
with the density sector. The $n$-th order propagator
$\varGamma^{(n)}_{\rm m}(\bm{k}_1,\ldots,\bm{k}_n)$ of the density
sector is defined by an ensemble average of the functional derivative:
\begin{equation}
  \left\langle
      \frac{\delta^n \delta_{\rm m}(\bm{k})}
      {\delta\delta_{\rm L}(\bm{k}_1)
        \cdots\delta\delta_{\rm L}(\bm{k}_n)}
  \right\rangle =
  (2\pi)^{3-3n}\delta_{\rm D}^3(\bm{k} - \bm{k}_{1\cdots n})
  \varGamma^{(n)}_{\rm m}(\bm{k}_1,\ldots,\bm{k}_n).
\label{eq:3-1}
\end{equation}
The original multi-point propagator $\varGamma^{(n)}_{ab_1\cdots b_n}$
with density and velocity sectors, defined in
Refs.~\cite{BCS08,BCS10}, is related to our definition of density
propagator by $\varGamma^{(n)}_{\rm m} = n! 2^{-n} (D_{\rm init}/D)^n
\varGamma^{(n)}_{1a_1\cdots a_n}u_{a_1}\cdots u_{a_n}$, where $D_{\rm
  init}$ is the linear growth factor at the initial time $t_{\rm
  init}$. The appearance of the Dirac's delta function on the
right-hand side of Eq.~(\ref{eq:3-1}) is due to the translational
symmetry. When the initial density field is random Gaussian, the
multi-point propagator of Eq.~(\ref{eq:3-1}) corresponds to the
coefficient of orthogonal expansion by a series of generalized
Wiener-Hermite functionals \cite{mat95}.

Substituting Eq.~(\ref{eq:1-3}) into the left-hand side of
Eq.~(\ref{eq:3-1}), we have
\begin{multline}
  \left\langle
      \frac{\delta^n \delta_{\rm m}(\bm{k})}
      {\delta\delta_{\rm L}(\bm{k}_1)
        \cdots\delta\delta_{\rm L}(\bm{k}_n)}
  \right\rangle = \frac{1}{(2\pi)^{3n}}
  \sum_{m=0}^\infty \frac{1}{m!}
  \int \frac{d^3k_1'}{(2\pi)^3}\cdots \frac{d^3k_m'}{(2\pi)^3}
\\ \times
  (2\pi)^3\delta_{\rm D}^3
  (\bm{k}-\bm{k}_{1\cdots n}-\bm{k}'_{1\cdots m})
  F_{n+m}(\bm{k}_1,\ldots\bm{k}_n,\bm{k}'_1,\ldots\bm{k}'_m)
\\ \times
  \left\langle \delta_{\rm L}(\bm{k}'_1)\cdots \delta_{\rm L}(\bm{k}'_m)
  \right\rangle.
\label{eq:3-2}
\end{multline}
The last factor is proportional to $\delta_{\rm
  D}^3(\bm{k}'_{1\cdots m})$ for the translational symmetry, and the
remaining factor is further decomposed into products of connected
polyspectra, $\sum \prod P^{(N)}$. For example,
\begin{align}
    \left\langle \delta_1 \delta_2 \delta_3 \delta_4
    \right\rangle &= 
    \left\langle \delta_1 \delta_2 \delta_3 \delta_4
    \right\rangle_{\rm c} +
    \left\langle \delta_1 \delta_2 \right\rangle_{\rm c}
    \left\langle \delta_3 \delta_4 \right\rangle_{\rm c}
\nonumber\\
    &\qquad +
    \left\langle \delta_1 \delta_3 \right\rangle_{\rm c}
    \left\langle \delta_2 \delta_4 \right\rangle_{\rm c} +
    \left\langle \delta_1 \delta_4 \right\rangle_{\rm c}
    \left\langle \delta_2 \delta_3 \right\rangle_{\rm c}
\nonumber\\
    &= \delta^{\rm D}_{1234} P^{(4)}_{1234}
    + \delta^{\rm D}_{12} \delta^{\rm D}_{34} P^{(2)}_{12} P^{(2)}_{34}
\nonumber\\
    &\qquad
    + \delta^{\rm D}_{13} \delta^{\rm D}_{24} P^{(2)}_{13} P^{(2)}_{24}
    + \delta^{\rm D}_{14} \delta^{\rm D}_{23} P^{(2)}_{14} P^{(2)}_{23},
\label{eq:3-3}
\end{align}
and so forth, where $\delta_1 = \delta(\bm{k}'_1)$, $\delta_2 =
\delta(\bm{k}'_2)$, $\delta^{\rm D}_{1234} = (2\pi)^3\delta_{\rm
  D}^3(\bm{k}'_{1234})$, $P^{(4)}_{1234} =
P(\bm{k}'_1,\ldots,\bm{k}'_4)$, etc. Thus we have
\begin{multline}
  \varGamma^{(n)}_{\rm m}(\bm{k}_1,\ldots,\bm{k}_n) = 
  \sum_{m=0}^\infty \frac{1}{m!}
  \int \frac{d^3k_1'}{(2\pi)^3}\cdots \frac{d^3k_m'}{(2\pi)^3}
\\ \times
  F_{n+m}(\bm{k}_1,\ldots,\bm{k}_n,\bm{k}'_1,\ldots,\bm{k}'_m)
  \left\langle \delta_{\rm L}(\bm{k}'_1)\cdots \delta_{\rm L}(\bm{k}'_m)
  \right\rangle.
\label{eq:3-4}
\end{multline}
This equation is equivalent to Eq.~(23) of Ref.~\cite{BCS10} when the
last term is decomposed into connected parts. The graphical
representation of Eq.~(\ref{eq:3-4}) is similar to Fig.~2 of
Ref.~\cite{BCS08}.

\begin{figure*}
\begin{center}
\includegraphics[width=42pc]{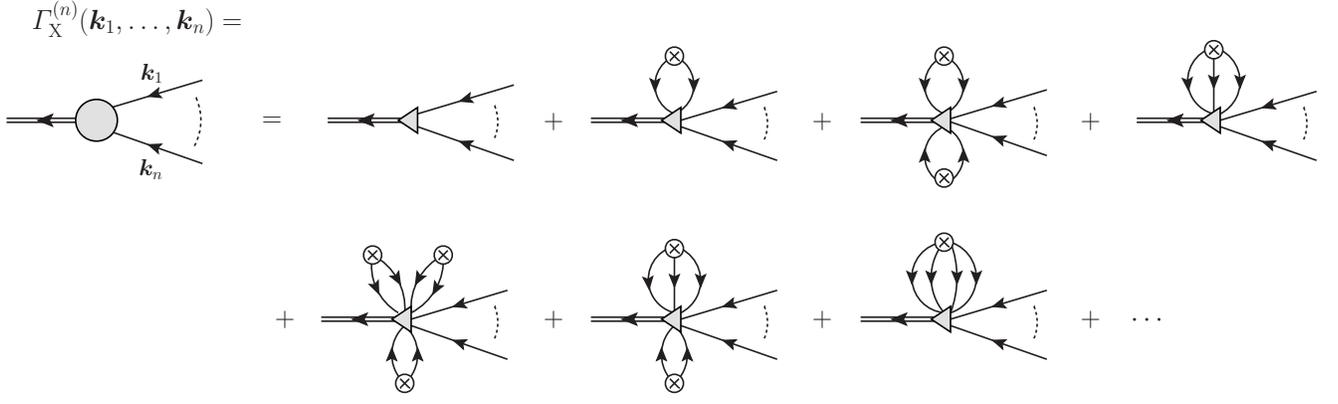}%{MultiP.eps}
\caption{\label{fig:MultiP} Diagrammatic representation of the
  multi-point propagator of objects in real/redshift space.}
\end{center}
\end{figure*}

Let us extend the above multi-point propagator of mass density field
to include the effects of biasing and redshift-space distortions. We
define the $n$-th order propagator $\varGamma^{(n)}_{\rm
  X}(\bm{k}_1,\ldots,\bm{k}_n)$ of the objects X by
\begin{equation}
  \left\langle
      \frac{\delta^n \delta_{\rm X}(\bm{k})}
      {\delta\delta_{\rm L}(\bm{k}_1)
        \cdots\delta\delta_{\rm L}(\bm{k}_n)}
  \right\rangle =
  (2\pi)^{3-3n}\delta_{\rm D}^3(\bm{k} - \bm{k}_{1\cdots n})
  \varGamma^{(n)}_{\rm X}(\bm{k}_1,\ldots,\bm{k}_n).
\label{eq:3-5}
\end{equation}
Substituting Eq.~(\ref{eq:1-5}) into the above equation, and following
the same way of obtaining the Eq.~(\ref{eq:3-4}), we have
\begin{multline}
  \varGamma^{(n)}_{\rm X}(\bm{k}_1,\ldots,\bm{k}_n) = 
  \sum_{m=0}^\infty \frac{1}{m!}
  \int \frac{d^3k_1'}{(2\pi)^3}\cdots \frac{d^3k_m'}{(2\pi)^3}
\\ \times
  K_{n+m}(\bm{k}_1,\ldots\bm{k}_n,\bm{k}'_1,\ldots\bm{k}'_m)
  \left\langle \delta_{\rm L}(\bm{k}'_1)\cdots \delta_{\rm L}(\bm{k}'_m)
  \right\rangle.
\label{eq:3-6}
\end{multline}
To include the redshift-space distortions, one simply apply the
substitution $K_{n+m} \rightarrow S_{n+m}$ in the above equation. The
diagrammatic representation of this equation is given in
Fig.~\ref{fig:MultiP}.

\begin{figure}
\begin{center}
\includegraphics[width=18pc]{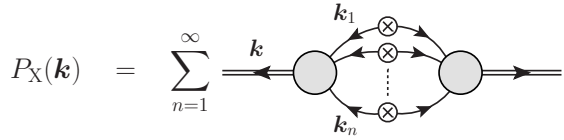}%{Pspec.eps}
\caption{\label{fig:Pspec} Diagrammatic representation of the power
  spectrum with multi-point propagators when the initial density field
  is Gaussian. Effects of biasing and redshift-space distortions are
  included in the multi-point propagators in our formalism.}
\end{center}
\end{figure}

The usage of the multi-point propagator is parallel to the one in
Refs.~\cite{BCS08,BCS10}. The multi-point propagator corresponds to
the summation of all the loops which are attached to each external
vertex. Therefore, the polyspectra $P^{(N)}_{\rm X}$ of
Eq.~(\ref{eq:1-7}) are represented by using the multi-point
propagators and corresponding diagrams do not have any loop which is
attached to a single external vertex of the multi-point propagator.
For example, Fig.~\ref{fig:Pspec} represents the power spectrum of the
object for the Gaussian initial conditions.
The resulting power spectrum is given by
\begin{multline}
  P_{\rm X}(\bm{k}) = \sum_{n=0}^\infty \frac{1}{n!}
  \int \frac{d^3k_1}{(2\pi)^3} \cdots \frac{d^3k_n}{(2\pi)^3}
  (2\pi)^3 \delta_{\rm D}^3(\bm{k} - \bm{k}_{1\cdots n})
\\ \times
  \left|\varGamma^{(n)}_{\rm X}(\bm{k}_1,\ldots,\bm{k}_n)\right|^2
  P_{\rm L}(k_1)\cdots P_{\rm L}(k_n).
\label{eq:3-7}
\end{multline}
As described in Ref.~\cite{BCS08}, it is important to note that each term
in the sum is positive and the subsequent contributions add
constructively.

\subsection{\label{subsec:LagResum} Lagrangian vertex resummations}

The multi-point propagators are still difficult to be exactly evaluated. One
of the remarkable results in the RPT is a derivation of the
propagators in the high-$k$ limit \cite{CS06a,CS06b,BCS08,BCS10}. It
is not trivial whether or not the same limit can be calculated in the
presence of bias. Investigations on this line will be interesting for
future work, and we leave them as an open question. Instead of
evaluating the full expression of the multi-point propagators, we
consider partial resummations of external vertex in the Lagrangian
representation of the perturbation theory in this section.

\begin{figure}
\begin{center}
\includegraphics[width=20pc]{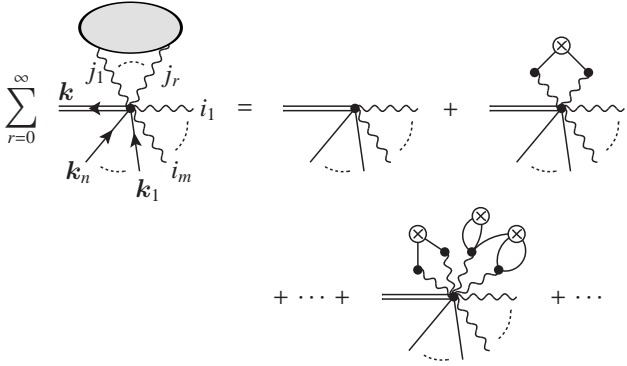}%{ResumL.eps}
\caption{\label{fig:ResumL} Vertex resummation with displacement legs.
}
\end{center}
\end{figure}

As obviously seen in Fig.~\ref{fig:MultiP}, the multi-point propagator
is essentially a renormalized external vertex. First we consider
partial resummation of the external vertex given in
Fig.~\ref{fig:ResumL}.
In this figure, the gray ellipse represents all the possible graphs
which are attached to a single external vertex with $r$ wavy lines.
The graphs in the gray ellipse are not necessarily connected, and can
be disconnected as illustrated in the second line of the figure.

According to the diagrammatic rules of Fig.~\ref{fig:LPTdiag} and
appropriate statistical factors, the corresponding factor of
Fig.~\ref{fig:ResumL} reduces to
\begin{align}
&
  \sum_{r=0}^\infty \frac{(-i)^r}{r!}
  \int \frac{d^3k'_1}{(2\pi)^3}\cdots \frac{d^3k'_r}{(2\pi)^3}
  \left\langle
      [\tilde{\varPsi}_{j_1}(\bm{k}'_1)] \cdots
      [\tilde{\varPsi}_{j_r}(\bm{k}'_r)] 
  \right\rangle
\nonumber\\
& \hspace{7pc}
  \times
  b^{\rm L}_n(\bm{k}_1,\ldots,\bm{k}_n)
  k_{i_1}\cdots k_{i_m}
  k_{j_1}\cdots k_{j_r}
\nonumber\\
& \quad = 
  \sum_{r=0}^\infty \frac{(-i)^r}{r!}
  \left\langle
      \left[
          \int\frac{d^3k'}{(2\pi)^3}
          \bm{k}\cdot\tilde{\bm{\varPsi}}(\bm{k}')
      \right]^r
  \right\rangle
  b^{\rm L}_n(\bm{k}_1,\ldots,\bm{k}_n)
  k_{i_1}\cdots k_{i_m}
\nonumber\\
& \quad = 
   \left\langle e^{-i\bm{k}\cdot\bm{\varPsi}} \right\rangle
  b^{\rm L}_n(\bm{k}_1,\ldots,\bm{k}_n)
  k_{i_1}\cdots k_{i_m},
\label{eq:3-30}
\end{align}
where 
\begin{equation}
   \bm{\varPsi} = \bm{\varPsi}(\bm{0})
   = \int\frac{d^3k'}{(2\pi)^3} \tilde{\bm{\varPsi}}(\bm{k}'),
\label{eq:3-31}
\end{equation}
is the displacement vector at the origin. The factor
\begin{equation}
   \varPi(\bm{k}) \equiv
   \left\langle e^{-i\bm{k}\cdot\bm{\varPsi}} \right\rangle
   = \int d^3\varPsi\, e^{-i\bm{k}\cdot\bm{\varPsi}} {\cal P}(\bm{\varPsi}),
\label{eq:3-32}
\end{equation}
is the characteristic function of the one-point distribution of the
displacement field, where ${\cal P}(\bm{\varPsi})$ is the one-point
probability function of the displacement field. This characteristic
function is a generating function of moments of $\bm{\varPsi}$ at a
single point in configuration space:
\begin{equation}
   \left\langle \varPsi_{j_1} \cdots \varPsi_{j_n} \right\rangle
   = \left. \frac{ i^n \partial^n \varPi(\bm{k})}
       {\partial k_{j_1}\cdots\partial k_{j_n}}\right|_{\bm{k}=\bm{0}}.
\label{eq:3-33}
\end{equation}
The characteristic function is represented by a connected moments by
the cumulant expansion theorem \cite{ma85}:
\begin{align}
   \varPi(\bm{k})
   & = \exp \left[ \sum_{n=1}^\infty \frac{(-i)^n}{n!}
       \left\langle (\bm{k}\cdot\bm{\varPsi})^n
       \right\rangle_{\rm c}
   \right]
\nonumber\\
   & = \exp \left[ \sum_{n=1}^\infty \frac{(-i)^n}{n!}
       k_{j_1}\cdots k_{j_n}
       \left\langle \varPsi_{j_1} \cdots \varPsi_{j_n}
       \right\rangle_{\rm c}
   \right].
\label{eq:3-34}
\end{align}

In real space, the cumulants of the last line of the above equation
are nonzero only when $n$ is an even number for the rotational
symmetry, and have the form
\begin{equation}
   \left\langle \varPsi_{j_1} \cdots \varPsi_{j_{2n}}
   \right\rangle_{\rm c}
   = A_{2n} \frac{2^n n!}{(2n)!}
     \left( \delta_{j_1j_2}\delta_{j_3j_4} \cdots\delta_{j_{2n-1}j_{2n}}
       + \mbox{perm.}
   \right),
\label{eq:3-35}
\end{equation}
where the factor $(2n)!/2^n n!$ is the number of all the possible
pairings among indices $j_1,\ldots j_{2n}$, and equals to the number
of terms in the parenthesis. When all $j_1,\ldots,j_{2n}$ take the
same component, e.g., $z$-axis, we have
\begin{align}
    \left\langle (\varPsi_z)^{2n} \right\rangle
    &= \left\langle|\bm{\varPsi}|^{2n} (\cos\theta)^{2n} \right\rangle_{\rm c} 
    = \left\langle|\bm{\varPsi}|^{2n}\right\rangle_{\rm c}
    \left\langle (\cos\theta)^{2n} \right\rangle
\nonumber\\
    &=  \frac{\langle |\bm{\varPsi}|^{2n} \rangle_{\rm c}}{2n+1},
\label{eq:3-36}
\end{align}
where $\theta$ is the polar angle of $\bm{\varPsi}$, and we used the
fact that the variables $|\bm{\varPsi}|$ and $\theta$ are independent
and the directional cosine is randomly distributed in real space.

Comparing Eqs.~(\ref{eq:3-35}) and (\ref{eq:3-36}), we have
\begin{equation}
   A_{2n} = \frac{\langle |\bm{\varPsi}|^{2n} \rangle_{\rm c}}{2n+1}.
\label{eq:3-36-1}
\end{equation}
This equation can also be directly confirmed by contracting
Eq.~(\ref{eq:3-35}) for each $n$. Thus Eq.~(\ref{eq:3-34}) reduces
to
\begin{equation}
   \varPi(k)
   = \exp\left[\sum_{n=1}^\infty \frac{(-1)^n A_{2n} k^{2n}}{(2n)!}
   \right]
   = \exp\left[\sum_{n=1}^\infty
     \frac{(-1)^n \langle |\bm{\varPsi}|^{2n} \rangle_{\rm c}}{(2n+1) (2n)!}
     k^{2n}
   \right].
\label{eq:3-37}
\end{equation}
When the higher-order cumulants of displacement field are negligible
on large scales, the above equation reduces to a Gaussian damping
factor,
\begin{equation}
  \varPi(k) \simeq
  \exp\left(-\frac{k^2}{6}\langle|\bm{\varPsi}|^2\rangle\right).
\label{eq:3-38}
\end{equation}
Therefore the large-scale power spectrum is smeared by nonlinear
effects, and such smearing is important in the analysis of BAO
\cite{CS08,mat08a}.

The above Eq.~(\ref{eq:3-37}), however, is valid only in real space
where the clustering is statistically isotropic. In redshift space,
the clustering is not statistically isotropic and Eq.~(\ref{eq:3-34})
should be evaluated with the displacement field $\bm{\varPsi}^{\rm s}$
of Eq.~(\ref{eq:2-51}). When the lowest-order (Zel'dovich)
approximation is valid on large scales, and higher-order cumulants of
the displacement field is negligible (i.e., primordial non-Gaussianity
is weak), we have again a Gaussian damping factor,
\begin{equation}
    \varPi(k) \simeq
    \exp\left\{-\frac{k^2}{6}
        \left[1+f(f+2)\mu^2\right]\langle|\bm{\varPsi}|^2\rangle
    \right\},
\label{eq:3-39}
\end{equation}
where
$\mu = k_z/k$ is the direction cosine of the wavevector with respect
to the lines of sight. This damping factor represents both effects of
nonlinear smearing and FoG in redshift space \cite{mat08a}, which are
present even on large scales. The damping factor of FoG is similar to,
but somewhat different from that adopted in a phenomenological
modeling \cite{PD94,sco04,tar10}, $\exp(-f^2\mu^2k^2\sigma_v^2/2)$,
where $\sigma_v^2$ is equal to $\langle|\bm{\varPsi}|^2\rangle/3$ at
the linear order.

\begin{figure}
\begin{center}
\includegraphics[width=20pc]{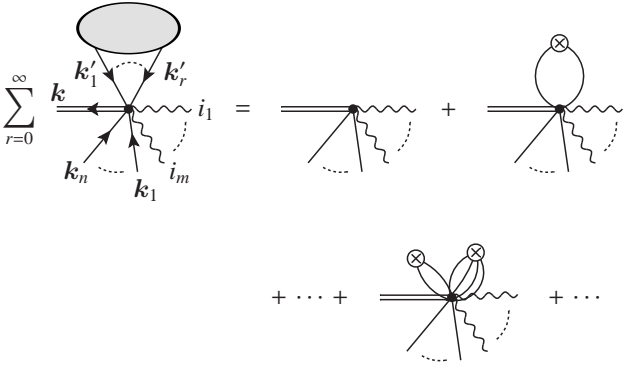}%{ResumB.eps}
\caption{\label{fig:ResumB} Vertex resummation with bias legs.
}
\end{center}
\end{figure}

Next we consider partial resummation of the external vertex given in
Fig.~\ref{fig:ResumB}. As in Fig.~\ref{fig:ResumL}, the gray ellipse
represents all the possible graphs which are attached to a single
external vertex with $r$ solid lines. The graphs in the ellipse are
not necessarily connected. The corresponding factor of
Fig.~\ref{fig:ResumB} reduces to
\begin{align}
&
  \sum_{r=0}^\infty \frac{1}{r!}
  \int \frac{d^3k'_1}{(2\pi)^3}\cdots \frac{d^3k'_r}{(2\pi)^3}
  b^{\rm L}_{n+r}(\bm{k}_1,\ldots,\bm{k}_n,\bm{k}'_1,\ldots,\bm{k}'_r)
\nonumber\\
& \hspace{9pc}
  \times
  \left\langle
      \delta_{\rm L}(\bm{k}'_1)\cdots\delta_{\rm L}(\bm{k}'_r)
  \right\rangle
  k_{i_1}\cdots k_{i_m}
\nonumber\\
& \quad =
  (2\pi)^{3n} \int \frac{d^3k'}{(2\pi)^3}
  \left\langle
      \frac{\delta^n\delta_{\rm X}^{\rm L}(\bm{k}')}
      {\delta\delta_{\rm L}(\bm{k}_1)\cdots\delta\delta_{\rm
          L}(\bm{k}_n)}
  \right\rangle k_{i_1}\cdots k_{i_m}
\nonumber\\
& \quad = c^{\rm L}_n(\bm{k}_1,\ldots,\bm{k}_n) k_{i_1}\cdots k_{i_m},
\label{eq:3-40}
\end{align}
where we define the renormalized nonlocal bias function in
Lagrangian space:
\begin{equation}
    c^{\rm L}_n(\bm{k}_1,\ldots,\bm{k}_n) = 
    (2\pi)^{3n} \int \frac{d^3k'}{(2\pi)^3}
    \left\langle
        \frac{\delta^n\delta_{\rm X}^{\rm L}(\bm{k}')}
        {\delta\delta_{\rm L}(\bm{k}_1)\cdots\delta\delta_{\rm
            L}(\bm{k}_n)}
    \right\rangle.
\label{eq:3-41}
\end{equation}
This expression is contrasted with Eq.~(\ref{eq:1-106-1}). Instead of
evaluating the functional derivatives at $\delta_{\rm L}=0$, taking
statistical averages of them gives the renormalized bias functions.

While the original function $b^{\rm L}_n$ can be determined solely by
a functional relation between biased field and linear density field in
Lagrangian space, the renormalized function $c^{\rm L}_n$ depends on
the statistical properties of the linear (initial) density field. In
configuration space, Eq.~(\ref{eq:3-41}) is equivalent to
\begin{equation}
    c^{\rm L}_n(\bm{q}-\bm{q}_1,\ldots,\bm{q}-\bm{q}_n) = 
    \left\langle
        \frac{\delta^n\delta_{\rm X}^{\rm L}(\bm{q})}
        {\delta\delta_{\rm L}(\bm{q}_1)\cdots\delta\delta_{\rm
            L}(\bm{q}_n)}
    \right\rangle,
\label{eq:3-42}
\end{equation}
where the translational invariance is taken into account. We use the
same symbols for variables both in Fourier space and configuration
space as long as the notation is obvious. When a model of the
Lagrangian bias is provided in configuration space, the renormalized
bias functions are evaluated by the above equation.

Taylor expansions are possible only when the number density field
$\delta^{\rm L}_{\rm X}$ is a smooth functional of the linear density
field $\delta_{\rm L}$. However, the renormalized bias function
$c^{\rm L}_n$ can be evaluated even when the biased field $\delta^{\rm
  L}_X$ is not a smooth functional of $\delta_{\rm L}$ and does not
have a Taylor expansion, since statistical average in
Eq.~(\ref{eq:3-41}) is possible even when the functional derivative in
the bracket is a singular functional. We will see some examples below
in which the bias cannot be expanded by a Taylor series while the
renormalized bias functions are still well-defined.

\begin{figure}
\begin{center}
\includegraphics[width=20pc]{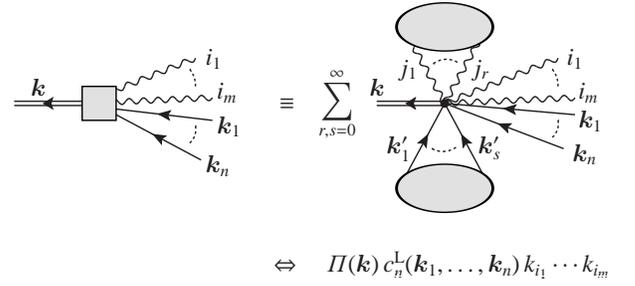}%{ResumLB.eps}
\caption{\label{fig:ResumLB} Partial resummation of the vertex.}
\end{center}
\end{figure}

\begin{figure}
\begin{center}
\includegraphics[width=8pc]{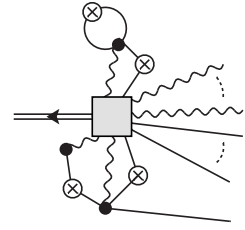}%{NonResumEx.eps}
\caption{\label{fig:NonResumEx} An example of graphs that are not
  resummed in Eq.~(\ref{eq:3-43}) or Fig.~\ref{fig:ResumLB}. }
\end{center}
\end{figure}

\begin{figure}
\begin{center}
\includegraphics[width=20pc]{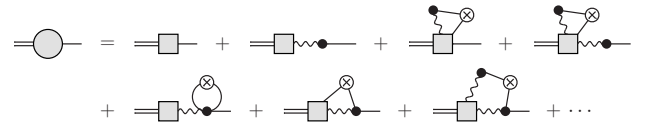}%{PropOne.eps}
\caption{\label{fig:PropOne} First-order propagator with partially
  resumed vertex up to one-loop contributions. }
\end{center}
\end{figure}

Putting the diagrams of Fig.~\ref{fig:ResumL} and
Fig.~\ref{fig:ResumB} together, we define the partial resummation of
the external vertex in Fig.~\ref{fig:ResumLB}.
The gray box in this figure represents the partial resummation with
all the sub-graphs which are attached to an external vertex with only
wavy lines and with only solid lines. The partial resummation of
Fig.~\ref{fig:ResumLB} results in the factor
\begin{equation}
   \varPi(\bm{k}) c^{\rm L}_n(\bm{k}_1,\ldots,\bm{k}_n)
   k_{i_1} \cdots k_{i_m}.
\label{eq:3-43}
\end{equation}
Connected graphs with both wavy and solid lines attached to an
external vertex are missed in this resummation. For example, graphs
like Fig.~\ref{fig:NonResumEx} are not included in the factor of
Eq.~(\ref{eq:3-43}).
To obtain the full multi-point propagators, all kinds of graphs like
the one in Fig.~\ref{fig:NonResumEx} should be added. For example, the
first-order propagator $\varGamma^{(2)}_{\rm X}$ is diagrammatically
given by Fig.~\ref{fig:PropOne} up to one-loop contributions.

\section{\label{sec:LBiasEx} Some models of the Lagrangian bias
}

\subsection{\label{subsec:LocLBias} Local Lagrangian bias
}

In the case of the local Lagrangian bias, the linear density field in
Lagrangian space, $\delta^{\rm L}_{\rm X}(\bm{q})$ is given by a
single function of the smoothed mass density field $\delta_R(\bm{q})$
at the same position, where $R$ is a smoothing radius. In
configuration space, we have
\begin{equation}
  \delta^{\rm L}_{\rm X} = F_{\rm X}\left(\delta_R\right),
\label{eq:4-1}
\end{equation}
where $F_{\rm X}$ is generally a nonlinear, univariate function, and
the smoothed mass density field in Lagrangian space is given by
\begin{equation}
  \delta_R(\bm{q}) = \int d^3q' W_R(|\bm{q}-\bm{q}'|)
  \delta_{\rm L}(\bm{q}'),
\label{eq:4-2}
\end{equation}
where the window function $W_R$ is spherically symmetric. Applying the
Taylor expansion and the Fourier transform to Eq.~(\ref{eq:4-1}),
the Lagrangian bias functions $b^{\rm L}_n$ in Eq.~(\ref{eq:1-106-1})
reduces to
\begin{equation}
  b^{\rm L}_n(\bm{k}_1,\ldots,\bm{k}_n)
  =  F_{\rm X}^{(n)}(0) W(k_1R)\cdots W(k_nR),
\label{eq:4-3}
\end{equation}
where $F_{\rm X}^{(n)} = d^n F_{\rm X}/d{\delta_R}^n$ is the $n$-th
derivative of the function $F_{\rm X}$, and $W(kR)$ is a
(3-dimensional) Fourier transform of the smoothing window function
$W_R(x)$. The renormalized bias function $c^{\rm L}_n$ of
Eq.~(\ref{eq:3-41}) reduces to
\begin{equation}
  c^{\rm L}_n(\bm{k}_1,\ldots,\bm{k}_n)
  = \left\langle F_{\rm X}^{(n)}(\delta_R) \right\rangle
  W(k_1R)\cdots W(k_nR).
\label{eq:4-4}
\end{equation}
By denoting ${\cal P}(\delta_R)$ as the probability distribution
function of $\delta_R$, the first factor of the above equation is
given by
\begin{align}
    \left\langle F_{\rm X}^{(n)}(\delta_R) \right\rangle
    &= \int_{-\infty}^\infty d\delta_R\, {\cal P}(\delta_R)\,
    F_{\rm X}^{(n)}(\delta_R)
\nonumber\\
    &= (-1)^n \int_{-\infty}^\infty d\delta_R\, {\cal P}^{(n)}(\delta_R)\,
    F_{\rm X}(\delta_R),
\label{eq:4-5}
\end{align}
where ${\cal P}^{(n)} = d^n{\cal P}/d{\delta_R}^n$.

On scales which are larger than the smoothing scale, $|\bm{k}_i| <
1/R$ ($i=1,\ldots,n$), one can ignore the window function $W(kR)$ in
Eqs.~(\ref{eq:4-3}) and (\ref{eq:4-4}). In this case the Lagrangian
bias functions are constants:
\begin{align}
  b^{\rm L}_n(\bm{k}_1,\ldots,\bm{k}_n) &\simeq F_{\rm X}^{(n)}(0),
\label{eq:4-5-1a}\\
  c^{\rm L}_n(\bm{k}_1,\ldots,\bm{k}_n) &\simeq
  \left\langle F_{\rm X}^{(n)}\right\rangle. 
\label{eq:4-5-1b}
\end{align}

One notices that Eq.~(\ref{eq:4-3}) is well-defined only when the bias
function $F_{\rm X}$ is a smooth function, because the factor $F_{\rm
  X}^{(n)}(0)$ corresponds to a coefficient of the Taylor series.
However, Eq.~(\ref{eq:4-4}) is applicable even when the bias function
is not a smooth function. As an illustration, we consider a
threshold bias given by
\begin{equation}
    F_{\rm X}(\delta_R)
    = C \varTheta(\delta_R - \delta_{\rm t}) - 1,
\label{eq:4-6}
\end{equation}
where a constant $\delta_{\rm t}$ is a threshold value,
\begin{equation}
   C = {\langle \varTheta(\delta_R - \delta_{\rm t}) \rangle}^{-1}
   = \left[ \int_{\delta_{\rm t}}^\infty d\delta_R\,
        \cal P(\delta_R)\right]^{-1},
\label{eq:4-7}
\end{equation}
and $\varTheta$ is the step function.
The Taylor expansion of this function is not well-defined since all
the derivatives at the origin are zero, $F_{\rm X}^{(n)}(0) = 0$
($n\geq 1$), and we have $b^{\rm L}_n = 0$ from Eq.~(\ref{eq:4-3}).
However, the expectation value in Eq.~(\ref{eq:4-4}) is not zero in
this case, and using Eq.~(\ref{eq:4-5}) we have
\begin{equation}
    \left\langle F_{\rm X}^{(n)}(\delta_R) \right\rangle =
        (-1)^n C \int_{\delta_{\rm t}}^\infty d\delta_R\,
        {\cal P}^{(n)}(\delta_R),
\label{eq:4-8}
\end{equation}
for $n\geq 1$. Therefore, the concept of renormalized bias functions
$c^{\rm L}_n$ extends the applicability to the case when the simple
Taylor expansion of the bias in Eq.~(\ref{eq:1-106}) does not work.

In the limit $\delta_{\rm t} \rightarrow \infty$, the threshold bias
can be considered as an approximation to the peaks bias in the
high-peak limit. Lagrangian statistics in this limit have been widely
studied in 1980's \cite{kai84,PW84,JS86,MLB86}. The methods developed
in those studies are essentially equivalent to applying the
renormalized bias in this work to the case of local Lagrangian bias.
The use of the partially resummed vertex of Eq.~(\ref{eq:3-43}) in
calculating loop-corrections to the power spectrum is equivalent to
applying a recent formalism developed in Ref.~\cite{mat08b} in a case
of local Lagrangian biasing and Gaussian initial conditions. The
present formalism is applicable even in cases of nonlocal Lagrangian
biasing with primordial non-Gaussianity.

\subsection{\label{subsec:MLBias} 
Multivariate Lagrangian bias
}

In a local bias model, the number density field $\delta^{\rm L}_{\rm
  X}$ is a function of the single variable $\delta_R$ of
Eq.~(\ref{eq:4-2}). We next consider the case in which the number
density field is a multivariate function of variables which are
convolutions of the linear density field. The multiple variables
$\chi_\alpha$, ($\alpha=1,2,\ldots$) in Lagrangian space are given by
\begin{equation}
  \chi_\alpha(\bm{q}) =
  \int d^3q' U_\alpha(\bm{q}-\bm{q}') \delta_{\rm L}(\bm{q}'),
\label{eq:4-50}
\end{equation}
where $U_\alpha$'s are the convolution kernels. For example, if one of
the variables $\chi_\alpha$ is the linear gravitational potential
$\varPhi_{\rm L}$ in Lagrangian space, the corresponding convolution
kernel is given by $U = -Ga^2\bar{\rho}_{\rm m}/|\bm{q}-\bm{q}'|$. If
one of the variables is a derivative of a smoothed density field, say
$\partial \delta_R(\bm{q})/\partial q_i$, the kernel is $U = \partial
W_R(|\bm{q}-\bm{q}'|)/\partial q_i$, and so forth.

The peaks bias is described by a function of the smoothed density
field and its derivatives up to second order,
$(\delta_R, \partial_i\delta_R, \partial_i\partial_j\delta_R)$
\cite{BBKS}. A multivariate bias model with two variables, $\delta_R$
and $\varPhi_{\rm L}$, is recently considered in the context of
scale-dependent halo bias with primordial non-Gaussianity
\cite{mcd08,GP10,BSS10}. A multivariate Eulerian bias model is also
proposed \cite{MR09}.

In general, we consider the biased field $\delta^{\rm L}_{\rm
  X}(\bm{q})$ is a local function of a finite number of the variables
$\chi_\alpha(\bm{q})$ at the same Lagrangian position. In
configuration space, we have
\begin{equation}
  \delta^{\rm L}_{\rm X}
  = F_{\rm X}\left(\chi_1,\chi_2,\ldots\right).
\label{eq:4-51}
\end{equation}
In this case, the Lagrangian bias function of Eq.~(\ref{eq:1-106})
reduces to
\begin{equation}
  b^{\rm L}_n(\bm{k}_1,\ldots,\bm{k}_n)
  = \sum_{\alpha_1,\ldots,\alpha_n} \left.\frac{\partial^n F_{\rm
      X}}{\partial\chi_{\alpha_1}\cdots\partial\chi_{\alpha_n}}
    \right|_{\chi_\alpha=0}
  U_{\alpha_1}(\bm{k}_1) \cdots U_{\alpha_n}(\bm{k}_n),
\label{eq:4-52}
\end{equation}
where $U_\alpha(\bm{k})$ is the Fourier transform of
$U_\alpha(\bm{q})$. The renormalized bias function of
Eq.~(\ref{eq:3-41}) reduces to
\begin{equation}
  c^{\rm L}_n(\bm{k}_1,\ldots,\bm{k}_n)
  = \sum_{\alpha_1,\ldots,\alpha_n} \left\langle\frac{\partial^n F_{\rm
      X}}{\partial\chi_{\alpha_1}\cdots\partial\chi_{\alpha_n}}
    \right\rangle
  U_{\alpha_1}(\bm{k}_1) \cdots U_{\alpha_n}(\bm{k}_n).
\label{eq:4-53}
\end{equation}

In the peaks model, for example, the variables $\chi_\alpha$
contain spatial derivatives of a smoothed density field,
$\partial\delta_R/\partial q_i$, $\partial^2 \delta_R/\partial
q_i\partial q_j$, and corresponding kernel windows are
$U_\alpha(\bm{k}) = ik_iW(kR), -k_ik_jW(kR)$, respectively. When one
of the variables $\chi_\alpha$ is given by the linear gravitational
potential $\varPhi_{\rm L}$, the kernel window $U_\alpha$ in Fourier
space is given by $U_\alpha(\bm{k}) = -4\pi G a^2 \bar{\rho}/k^2$.

\subsection{\label{subsec:HaloBias}
Bias functions from universal mass function
}

One of the most popular models of biasing in nonlinear structure
formation is provided by the halo approach
\cite{MW96,MJW97,ST99,SSHJ01,CS02}, which is based on the extended
Press-Schechter theory \cite{PS74,eps83,PH90,bow91,BCEK91}. The
peak-background split is applied in this approach, and the bias of
halos are considered as a local Lagrangian bias. The bias functions
can be calculated by Eq.~(\ref{eq:4-3}) or Eq.~(\ref{eq:4-4}) in this
case. In this subsection, we derive explicit expressions of the bias
functions of halos. The derivation is similar to that of
Ref.~\cite{mat08b}, in which the Gaussian initial conditions are
assumed, and unfortunately the powers of growth factor in final
expressions are incorrect. Below we correct the expressions of the
last reference and give a derivation which applies even when the
initial density field is non-Gaussian in general.

The mass of halo is related to the Lagrangian radius $R$ of a
spherical cell by $M = 4\pi \bar{\rho}R^3/3$, or $R =[M/(1.162\times
10^{12}h^{-1}M_\odot \varOmega_{\rm m})]^{1/3} \,h^{-1}{\rm Mpc}$,
where $M_\odot = 1.989\times 10^{30}\,{\rm kg}$ is the solar mass, and
$\varOmega_{\rm m}$ is the density parameter at the present time.
Henceforth, $\sigma^2(M)$ denotes the variance of density fluctuations
smoothed on a mass scale $M$ which is linearly extrapolated to the
present time.

According to the Press-Schechter theory and its extensions, the
comoving number density of halos with a mass range $dM$ around $M$,
identified at redshift $z$, is given by
\begin{equation}
  n(M,z)dM = \frac{\bar{\rho}}{M} f_{\rm MF}(\nu)\,d\ln\nu,
\label{eq:4-101}
\end{equation}
where $\nu = \delta_{\rm c}(z)/\sigma(M)$ is the typical amplitude of
fluctuations that produce those halos, $\delta_{\rm c}(z) =
\varDelta_{\rm c}/D(z)$, and $\varDelta_{\rm c}$ is the critical
overdensity for spherical collapse at the redshift $z$. In the
Einstein-de~Sitter model, the critical overdensity is independent on
redshift, $\varDelta_{\rm c} = 3(3\pi/2)^{2/3}/5 \simeq 1.686$, and
only weakly depends on cosmological parameters and redshift in general
cosmology. Since the condition of collapse is always expressed by the
linearly extrapolated overdensity at the present time, the growth
factor is absorbed into the critical overdensity $\delta_{\rm c}(z)$.
The multiplicity function $f_{\rm MF}(\nu)$ is normalized by
\begin{equation}
  \int_0^\infty f_{\rm MF}(\nu)\, \frac{d\nu}{\nu} = 1,
\label{eq:4-102}
\end{equation}
to ensure all the mass in the universe is contained in halos in the
limit $D(z) \rightarrow \infty$.

In the original Press-Schechter (PS) theory, the multiplicity function
$f_{\rm MF}(\nu)$ is given by
\begin{equation}
  f_{\rm PS}(\nu) = \sqrt{\frac{2}{\pi}}\,\nu e^{-\nu^2/2}.
\label{eq:4-103}
\end{equation}
The original PS mass function is improved by Sheth and Tormen (ST)
\cite{ST99} to give a better fit in numerical simulations of CDM-type
cosmologies with Gaussian initial conditions. The corresponding
multiplicity function is given by
\begin{equation}
  f_{\rm ST}(\nu) = A(p) \sqrt{\frac{2}{\pi}}
  \left[1 + \frac{1}{(q\nu^2)^p}\right]\sqrt{q}\,\nu e^{-q\nu^2/2},
\label{eq:4-104}
\end{equation}
where $p=0.3$, $q=0.707$ are numerically fitted parameters, and $A(p)
= [1+\pi^{-1/2}2^{-p}\varGamma(1/2-p)]^{-1}$ is the normalization
factor. The ST mass function is applicable only for Gaussian initial
conditions. When the non-Gaussianity is present in the initial density
field, the multiplicity function should have the correction factor
\cite{MVJ00,LV08}.

In the extended PS theory, the number density of halos of mass $M$,
identified at redshift $z$, in a region of Lagrangian radius $R_0$ in
which the linear overdensity extrapolated to the present time is
$\delta_0$, is given by \cite{CS02}
\begin{equation}
  n(M,z|\delta_0,R_0)dM = \frac{\bar{\rho}}{M} f_{\rm
    MF}(\nu')\, d\ln\nu',
\label{eq:4-105}
\end{equation}
where
\begin{equation}
  \nu' =
  \frac{\delta_{\rm c}(z) - \delta_0}
    {\left[\sigma^2(M) - \sigma_0^2\right]^{1/2}},\ 
    \sigma_0 = \sigma(M_0),\ M_0 = 4\pi\bar{\rho}R_0^3/3. 
\label{eq:4-106b}
\end{equation}
The halo of
mass $M$ is collapsed at $z$, while $M_0$ is assumed uncollapsed at
$z=0$, and thus we always have $\delta_{\rm c}(z) > \delta_0$. The
conditional number density of Eq.~(\ref{eq:4-105}) represents the
biasing for the Lagrangian number density of halos. The
smoothed density contrast $\delta_0$ of mass modulates the number of
halos. The density contrast of halos in Lagrangian space is given by
\begin{equation}
  \delta^{\rm L}_{\rm h} = \frac{n(M,z|\delta_0,R_0)}{n(M,z)}-1.
\label{eq:4-107}
\end{equation}

Since $d\ln\nu'/d\ln\nu = \sigma^2(M)/[\sigma^2(M)-\sigma_0^2]$, we
have
\begin{equation}
  \delta^{\rm L}_{\rm h} = \frac{\sigma^2(M)}{\sigma^2(M)-\sigma_0^2}
  \frac{f_{\rm MF}(\nu')}{f_{\rm MF}(\nu)}-1.
\label{eq:4-108}
\end{equation}
This relation gives the function $F_{\rm X}(\delta_R)$ of
Eq.~(\ref{eq:4-1}), where the smoothing radius in Eq.~(\ref{eq:4-1})
correspond to $R \rightarrow R_0$ here. We assume the redshift $z$ of
halo identification is the same as the redshift of halo observation.
The smoothed linear density field in Eq.~(\ref{eq:4-1}) corresponds to
$\delta_R \rightarrow D(z)\delta_0$ here, because $\delta_0$ is the
value extrapolated to the present time.

To evaluate the bias functions, the derivatives $F_{\rm X}^{(n)}$ in
Eqs.~(\ref{eq:4-3}), (\ref{eq:4-4}) need to be derived. We consider a
limit of the peak-background split for simplicity, and assume
$\sigma^2(M) \gg \sigma_0^2$ (However, see Ref.~\cite{MSS10} for
limitations of this commonly used method.). In this limit, we have
\begin{equation}
  F_{\rm X}^{(n)}(\delta_R) \simeq
  \frac{1}{D^n(z)}
  \left(\frac{\partial}{\partial {\delta_0}}\right)^n
  \delta^{\rm L}_{\rm h} = \left(\frac{-1}{D(z)\sigma(M)}\right)^n 
  \frac{f^{(n)}_{\rm MF}(\nu')}{f_{\rm MF}(\nu)},
\label{eq:4-109}
\end{equation}
for $n\geq 1$, where $f_{\rm MF}^{(n)}$ is the $n$-th derivative of
the multiplicity function $f_{\rm MF}$. The substitution $\delta_0 = 0$
is equivalent to $\nu' = \nu$ in the present limit. In the same limit,
taking the statistical average over the distribution of $\delta_0$
also equivalent to substituting $\nu'=\nu$ in Eq.~(\ref{eq:4-109}),
because the distribution function of $\delta_0$ is highly peaked at
$\delta_0=0$ and its variance is much smaller than $\delta_{\rm
  c}^2(z)$. Therefore, we have
\begin{equation}
  F_{\rm X}^{(n)}(0) \simeq
  \left\langle F_{\rm X}^{(n)}(\delta_R) \right\rangle
      \simeq
  \left(\frac{-1}{D(z)\sigma(M)}\right)^n 
  \frac{f^{(n)}_{\rm MF}(\nu)}{f_{\rm MF}(\nu)},
\label{eq:4-110}
\end{equation}
in the limit of peak-background split, $\sigma^2(M) \gg \sigma_0^2$.
The Lagrangian bias functions of Eqs.~(\ref{eq:4-3}), (\ref{eq:4-4})
reduces to
\begin{multline}
  b^{\rm L}_n(\bm{k}_1,\ldots,\bm{k}_n) =
  c^{\rm L}_n(\bm{k}_1,\ldots,\bm{k}_n)
  = \left(\frac{-1}{D(z)\sigma(M)}\right)^n 
  \frac{f^{(n)}_{\rm MF}(\nu)}{f_{\rm MF}(\nu)},
\label{eq:4-111}
\end{multline}
where the window function to define the background field $\delta_0$ is
dropped, assuming the large-scale limit $|\bm{k}_i| \ll 1/R_0$. The
last approximation is consistent with that of the peak-background
split. The above Eq.~(\ref{eq:4-111}) is applicable even in
non-Gaussian initial conditions, as long as effects of non-Gaussianity
are taken into account in the multiplicity function.

In the case of ST mass function of Eq.~(\ref{eq:4-104}) with Gaussian
initial conditions, the above bias functions are given by
\begin{align}
  b^{\rm L}_1 &= \frac{1}{\varDelta_{\rm c}}
  \left[q\nu^2 - 1 + \frac{2p}{1 + (q\nu^2)^p}\right],
\label{eq:4-112a}\\
  b^{\rm L}_2 &= \frac{1}{{\varDelta_{\rm c}}^2}
  \left[q^2\nu^4 - 3 q\nu^2 + \frac{2p(2q\nu^2 + 2p -1)}{1 +
        (q\nu^2)^p}\right],
\label{eq:4-112b}
\end{align}
and so forth. Essentially the same expressions are derived in
Ref.~\cite{mat08b}. Unfortunately, the factor $1/D^n(z)$ is
incorrectly missing in Eq.~(55) of Ref.~\cite{mat08b}. Accordingly,
the factor $\delta_{\rm c}(z)$ which appear in Eqs.~(57)--(59) of that
paper should be replaced by $\varDelta_{\rm c} = D(z)\delta_{\rm
  c}(z)$.

\subsection{\label{subsec:PeakBias} Some properties of peaks bias }

In the peaks formalism, a density peak in Lagrangian space is
considered as a location of structure formation \cite{BBKS}. In the
smoothed density field $\delta_R(\bm{q})$, the number density of peaks
above a height $\nu$ is given by
\begin{equation}
  n_{\rm pk}
  = \theta\left(\delta_R/\sigma_R - \nu\right)
  \delta_{\rm D}^3\left(\bm{\nabla}\delta_R\right)
  \left|\det\left(\bm{\nabla}\bm{\nabla}\delta_R\right)\right|
  \theta\left(\lambda_3\right)
\label{eq:4-150}
\end{equation}
where $\sigma_R=\langle\delta_R^2\rangle^{1/2}$, and $\lambda_3$ is
the smallest eigenvalue of the matrix $[-\bm{\nabla} \bm{\nabla}
\delta_R]_{ij} = -\partial_i \partial_j \delta_R$. Thus the number
density of peaks is a multivariate function of a scalar $\delta_R$, a
vector $\bm{\nabla}\delta_R$, and a tensor
$\bm{\nabla}\bm{\nabla}\delta_R$ at each position.

The number density of Eq.~(\ref{eq:4-150}) is a singular function.
Therefore the Taylor expansion cannot be applied. The unrenormalized
bias functions $b_n$ are not well-defined in this case, and it is
crucial to consider the renormalized bias functions of
Eq.~(\ref{eq:3-41}). The peaks bias is one of the multivariate
Lagrangian bias as described in Sec.~\ref{subsec:MLBias}, and the
renormalized bias function is given by Eq.~(\ref{eq:4-53}) where
$(\chi_\alpha) =
(\delta_R,\bm{\nabla}\delta_R,\bm{\nabla}\bm{\nabla}\delta_R)$ is a
10-dimensional vector. Since $\bm{\nabla}\bm{\nabla}\delta_R$ is a
symmetric tensor, only six components of
$\partial_i\partial_j\delta_R$ are independent. The corresponding
kernels in Eq.~(\ref{eq:4-50}) are given by $(U_\alpha)=[W(kR),
ik_iW(kR), -k_ik_jW(kR)]$ where $i\leq j$.

The calculation of the coefficient $\langle \partial^n F_{\rm
  X}/\partial\chi_{\alpha_1}\cdots\partial\chi_{\alpha_n}\rangle$ for
general $n$ in the peaks model is quite involved. In
Ref.~\cite{des08,DCSS10}, the correlation function and the power
spectrum of peaks are calculated up to second order in the case of
Gaussian initial condition. In this paper, we do not derive explicit
forms of the coefficients. Instead, we consider formal properties of
the bias functions derived from the rotational symmetry below.

For $n=1$, we have
\begin{multline}
  c_1^{\rm L}(\bm{k}) =
  W(kR)
  \left[
      \left\langle\frac{\partial F_{\rm X}}{\partial \delta_R}\right\rangle
    + i\sum_i k_i \left\langle\frac{\partial F_{\rm X}}
        {\partial \delta_{R,i}}\right\rangle
  \right. \\
  \left.
    - \sum_{i,j} k_i k_j
      \left\langle\frac{\partial F_{\rm X}}
        {\partial \delta_{R,ij}}\right\rangle
    \right],
\label{eq:4-151}
\end{multline}
where $\delta_{R,i} = \partial_i\delta_R$ and $\delta_{R,ij}
= \partial_i \partial_j\delta_R$. In the above equation, the function
$F_{\rm X}$ is symmetrized with respect to the off-diagonal
derivatives $\partial_i\partial_j\delta_R$, and partial derivatives
are taken as if $\partial_i\partial_j\delta_R$ and
$\partial_j\partial_i\delta_R$ were independent when $i\ne j$. From
the rotational symmetry, the second term in the square parenthesis in
Eq.~(\ref{eq:4-151}) identically vanishes. The last term is
proportional to $k^2$, since $\langle\partial F_{\rm
  X}/\partial\delta_{R,ij}\rangle \propto \delta_{ij}$. Thus, the
scale dependence of the first-order bias function should have a form,
\begin{equation}
  c_1^{\rm L}(\bm{k}) = W(kR)\left( A_1 + B_1 k^2 \right),
\label{eq:4-152}
\end{equation}
where
\begin{equation}
  A_1 = \left\langle\frac{\partial F_{\rm X}}{\partial \delta_R}\right\rangle,
  \quad
  B_1 = -\frac13 \sum_i
  \left\langle\frac{\partial F_{\rm X}}
      {\partial \delta_{R,ii}}\right\rangle.
\label{eq:4-153}
\end{equation}
This form is exact for peaks bias models \cite{mat99,DCSS10}, and
higher-order powers of $k^n$ with $n\geq 3$ do not appear.

It is easily understood that the first bias function of peaks should
have the form of Eq.~(\ref{eq:4-152}). The peaks are defined by up to
second derivatives of the smoothed field, and thus the first bias
functions of Eqs.~(\ref{eq:4-52}), (\ref{eq:4-53}) involve only
polynomials of wavevector $\bm{k}$ up to second order. Since the bias
function is rotationally invariant, only the form of
Eq.~(\ref{eq:4-152}) is allowed.

For $n=2$, the same considerations show that the bias function should
have a form,
\begin{multline}
  c_2^{\rm L}(\bm{k}_1,\bm{k}_2) =
  W(k_1R)W(k_2R)
  \left[
      A_2 + B_2 \left(k_1^2+k_2^2\right)
  \right.
\\
  \left.
    + C_2 \bm{k}_1\cdot\bm{k}_2 
    + D_2 k_1^2 k_2^2 + E_2
    \left(\bm{k}_1\cdot\bm{k}_2\right)^2
  \right],
\label{eq:4-154}
\end{multline}
where 
\begin{align}
  A_2 &= \left\langle\frac{\partial^2 F_{\rm X}}{\partial
        {\delta_R}^2}\right\rangle,
\label{eq:4-155a}\\
  B_2 &= -\frac13 \sum_i
    \left\langle\frac{\partial^2 F_{\rm X}}
      {\partial \delta_R \partial \delta_{R,ii}}\right\rangle,
\label{eq:4-155b}\\
  C_2 &= -\frac13 \sum_i
    \left\langle\frac{\partial^2 F_{\rm X}}
      {\partial {\delta_{R,i}}^2}\right\rangle,
\label{eq:4-155c}\\
  D_2 &= \frac{2}{15} \sum_{i,j}
    \left\langle\frac{\partial^2 F_{\rm X}}
      {\partial \delta_{R,ii}
        \partial \delta_{R,jj}}\right\rangle
  - \frac{1}{15} \sum_{i,j}
    \left\langle\frac{\partial^2 F_{\rm X}}
      {\partial {\delta_{R,ij}}^2}\right\rangle,
\label{eq:4-155d}\\
  E_2 &= -\frac{1}{30} \sum_{i,j}
    \left\langle\frac{\partial^2 F_{\rm X}}
      {\partial \delta_{R,ii}
        \partial \delta_{R,jj}}\right\rangle
  + \frac{1}{10} \sum_{i,j}
    \left\langle\frac{\partial^2 F_{\rm X}}
      {\partial {\delta_{R,ij}}^2}\right\rangle.
\label{eq:4-155e}
\end{align}
It is again easily understood that the second bias function of peaks
should have the form of Eq.~(\ref{eq:4-154}), since the function is
rotationally invariant and made from polynomials of wavevectors
$\bm{k}_1$ and $\bm{k}_2$ up to second order for each. The explicit
evaluations of the above coefficients are tedious. In
Ref.~\cite{DCSS10}, second-order biased correlation function with a
Gaussian initial condition is analytically calculated. Similar
techniques should be also useful in our formalism, which we leave for
future work.

\section{Conclusions
\label{sec:concl}
}

In this paper, the standard nonlinear perturbation theory of the
gravitational instability is extended in several directions. One of
the main extensions is the inclusion of the nonlocal bias, which is
a general framework of biasing. Nonlocal biases both in
Eulerian and Lagrangian spaces are formulated and consistently
included in EPT and LPT, respectively. The nonlinear Eulerian and
Lagrangian biases are compatible only in the framework of nonlocal
bias. The relations among perturbation kernels of EPT and LPT with
nonlocal biases are derived.

Effects of redshift-space distortions and primordial non-Gaussianity
are also included in our formalism. Therefore, our formalism provides
a complete theory to predict the observable quantities in redshift
surveys, once a model of bias and cosmology are given.

The concept of vertex resummations in the presence of nonlocal bias is
introduced. We show that the vertex resummation of the bias extends
the applicability of the formalism to the case when the bias
function(al) cannot be expanded into a Taylor series. This extension
is essential for handling, e.g., the threshold bias and the peaks
model, in which the bias involves singular functions such as the
Heaviside's step function, Dirac's delta function, etc. Calculation of
perturbative bias functions in our formalism is exemplified by
considering some models of local and nonlocal models of Lagrangian
bias, such as the threshold bias model, the multivariate bias model,
the halo model, and the peaks model. The scale dependence of bias
functions are straightforwardly obtained in our formalism once a model
of nonlocal bias is given. The scale dependence of Eulerian bias
arises both from the nonlocal Lagrangian bias and nonlinear
evolutions.

The formalism of the present paper provides a basic methodology for
future applications of the perturbation theory. For example, the
scale-dependent bias in the presence of primordial non-Gaussianity has
been derived by adopting either halo models \cite{dal08,MV08,slo08} or
the local Eulerian bias \cite{TKM08}. Our formalism allows to
calculate the scale-dependent bias in any models of bias in a
consistent manner \cite{mat11}. The power spectrum with BAO in any
given models of bias can be calculated with our formalism. The result
of applying our formalism to a local Lagrangian bias is equivalent to
the work in Ref.~\cite{mat08b}. More precise modeling of the bias
would be required in future analysis of the BAO in the galaxy power
spectrum to constrain the nature of dark energy.

Our formalism provides a way to perturbatively calculate the nonlinear
power spectrum, bispectrum, trispectrum, and other polyspectra. These
polyspectra are fundamental statistics and any other statistics in the
large-scale structure, such as the correlation functions,
counts-in-cells, genus statistics, etc.~are expressible by these
polyspectra, in principle. The formalism developed in this paper would
have an essential importance in the era of precision cosmology with
the large-scale structure of the universe.

\begin{acknowledgments}
    I wish to thank R.~Sheth for helpful discussion. I acknowledge
    support from the Ministry of Education, Culture, Sports, Science,
    and Technology, Grant-in-Aid for Scientific Research (C),
    21540263, 2009, and Grant-in-Aid for Scientific Research on
    Priority Areas No. 467 ``Probing the Dark Energy through an
    Extremely Wide and Deep Survey with Subaru Telescope.'' This work
    is supported in part by JSPS (Japan Society for Promotion of
    Science) Core-to-Core Program ``International Research Network for
    Dark Energy.''
\end{acknowledgments}

\bigskip
\appendix

\section{\label{app:SphericalModel}
Bias Parameters in the Spherical Collapse Model
}

In this Appendix, the relations between the Eulerian and Lagrangian
bias parameters in the spherical collapse model are derived, following
and generalizing the argument of Ref.~\cite{MJW97}. The
Einstein--de~Sitter universe is assumed in the following equations for
simplicity. The dependences of the results on cosmological parameters
are weak.

The time evolution of proper radius $r$ for a spherical mass shell as
a function of the scale factor $a$ is given by a parametric form
\cite{pee80}
\begin{align}
  \frac{r}{r_{\rm i}} &= \frac{3}{10} \frac{1-\cos\theta}{\delta_{\rm i}},
\label{eq:a-1a}\\
  \frac{a}{a_{\rm i}} &= \frac{3}{10}\left(\frac92\right)^{1/3}
  \frac{(\theta-\sin\theta)^{2/3}}{\delta_{\rm i}},
\label{eq:a-1b}
\end{align}
where $r_{\rm i}$ and $a_{\rm i}$ are initial values of $r$ and $a$,
respectively, and $\delta_{\rm i}$ is the initial density contrast at
$a_{\rm i}$. For $\delta_{\rm i}>0$, the parameter $\theta$ is a
positive real number. For $\delta_{\rm i} < 0$, the replacement
$\theta \rightarrow i\theta$ should be applied to have the parameter
real and positive.

The comoving radius is given by $R = r/a$, and therefore an
overdensity $\rho/\bar{\rho}$ of any kind in the spherical volume is
enhanced by a factor of $(R_{\rm i}/R)^3 = (a/a_{\rm i})^3(r_{\rm
  i}/r)^3 = (9/2)(\theta-\sin\theta)^2/(1-\cos\theta)^3$. Thus, the
density contrasts of mass $\delta_{\rm m}$ and of biased object
$\delta_{\rm X}$ in the spherical volume are given by $1+\delta_{\rm
  m} = (R_{\rm i}/R)^3$ in the limit of $|\delta_{\rm i}| \ll 1$, and
$1+\delta_{\rm X} = (R_{\rm i}/R)^3(1+\delta^{\rm L}_{\rm X})$, where
$\delta^{\rm L}_{\rm X}$ is the density contrast in Lagrangian space.
Thus, we have
\begin{align}
  1 + \delta_{\rm m} &=
  \frac92 \frac{(\theta-\sin\theta)^2}{(1-\cos\theta)^3},
\label{eq:a-2a}\\
  1 + \delta_{\rm X}  &=
  \left(1 + \delta_{\rm m}\right)
  \left(1 + \delta^{\rm L}_{\rm X}\right).
\label{eq:a-2b}
\end{align}
The form of Eq.~(\ref{eq:a-2a}) is well-known \cite{pee80}. The
Eq.~(\ref{eq:a-2b}) can also be derived from general
Eqs.~(\ref{eq:1-102}) and (\ref{eq:1-103}) in the case of spherical
perturbations.

The linear density contrast $\delta_{\rm L}$ is proportional to the
scale factor $a$, and from Eq.~(\ref{eq:a-1b}), we have
\begin{equation}
  \delta_{\rm L}
  = \frac{3}{10}\left(\frac92\right)^{1/3} (\theta-\sin\theta)^{2/3}.
\label{eq:a-3}
\end{equation}
For $|\delta_{\rm m}| \ll 1$, the relation between $\delta_{\rm m}$
and $\delta_{\rm L}$ is derived as a power series by Taylor expansions
of Eqs.~(\ref{eq:a-2a}) and (\ref{eq:a-3}) with respect to the
parameter $\theta$. The results are
\begin{align}
  \delta_{\rm m} &= \delta_{\rm L} + \frac{17}{21} {\delta_{\rm L}}^2
  + \frac{341}{567} {\delta_{\rm L}}^3 + \frac{55805}{130977}
  {\delta_{\rm L}}^4 + \cdots,
\label{eq:a-4a}\\
  \delta_{\rm L} &= \delta_{\rm m} - \frac{17}{21} {\delta_{\rm m}}^2
  + \frac{2815}{3969} {\delta_{\rm m}}^3 - \frac{590725}{916839}
  {\delta_{\rm m}}^4 + \cdots.
\label{eq:a-4b}
\end{align}
The Eq.~(\ref{eq:a-4a}) is derived in Ref.~\cite{ber92}. Although the
Eq.~(\ref{eq:a-4b}) is described in Refs.~\cite{MJW97,CS02}, they put
incorrect numbers in the coefficients of third- and fourth-order
terms. The coefficients $a_3$ and $a_4$ in their Eq.~(A4) of
Ref.~\cite{MJW97} should be replaced by $a_3=2815/3969$ and
$a_4=-590725/916389$.

The dynamical evolutions are local in the spherical collapse model. In
this special case, both the Eulerian and Lagrangian biases can be
simultaneously local. We have expansions
\begin{align}
  \delta_{\rm X} &= \sum_{n=1}^\infty \frac{b_n}{n!} {\delta_{\rm
      m}}^n,
  \label{eq:a-5a}\\
  \delta^{\rm L}_{\rm X} &= \sum_{n=1}^\infty \frac{b^{\rm L}_n}{n!}
  {\delta_{\rm L}}^n,
\label{eq:a-5b}
\end{align}
where $b_n$ and $b^{\rm L}_n$ are constant bias parameters. Putting
Eqs.~(\ref{eq:a-2b}), (\ref{eq:a-4b}), (\ref{eq:a-5b}) together, we
have a series expansion of $\delta_{\rm X}$ in terms of $\delta_{\rm
  m}$. Equating the resulting coefficients with Eq.~(\ref{eq:a-5a})
gives
\begin{align}
  b_1 &= b^{\rm L}_1+1,
\label{eq:a-6a}\\
  b_2 &= b^{\rm L}_2 + \frac{8}{21} b^{\rm L}_1,
\label{eq:a-6b}\\
  b_3 &= b^{\rm L}_3  - \frac{13}{7} b^{\rm L}_2 - \frac{796}{1323} b^{\rm L}_1,
\label{eq:a-6c}\\
  b_4 &= b^{\rm L}_4  - \frac{40}{7} b^{\rm L}_3 + \frac{7220}{1323}
  b^{\rm L}_2 + \frac{476320}{305613} b^{\rm L}_1.
\label{eq:a-6d}
\end{align}
Since the relation of Eq.~(\ref{eq:a-4b}) depends only very weakly on
cosmological model \cite{ber92}, the relations of
Eqs.~(\ref{eq:a-6a})--(\ref{eq:a-6d}) also do so.

%%%%%%%%%%%%
%\newcommand{\apj}{Astrophys. J.}
\newcommand{\aap}{Astron. Astrophys. }
\newcommand{\apjl}{Astrophys. J. Letters }
\newcommand{\apjs}{Astrophys. J. Suppl. Ser. }
\newcommand{\apss}{Astrophys. Space Sci. }
\newcommand{\jcap}{J. Cosmol. Astropart. Phys. }
\newcommand{\mnras}{Mon. Not. R. Astron. Soc. }
\newcommand{\mpla}{Mod. Phys. Lett. A }
\newcommand{\pasj}{Publ. Astron. Soc. Japan }
\newcommand{\physrep}{Phys. Rep. }
\newcommand{\ptp}{Progr. Theor. Phys. }
\newcommand{\jetp}{JETP }
%\newcommand{\prl}{Phys. Rev. Lett.}

%\bibliography{redoneloop}% Produces the bibliography via BibTeX.

\end{document}